\documentclass[12pt]{article}
\usepackage{booktabs}  
\usepackage{amsmath}
\usepackage{amsthm}
\usepackage{amssymb}
\usepackage{mathrsfs}
\usepackage{gensymb}
\usepackage{textcomp}  
\usepackage{xcolor}
\usepackage{lipsum}    
\usepackage{graphicx}
\usepackage[font=footnotesize]{caption}
\usepackage{subcaption}
\usepackage{bm, bbm}
\RequirePackage{diagbox, adjustbox}
\usepackage[colorlinks=true,linkcolor=black, citecolor=black, urlcolor=black]{hyperref}
\usepackage{multirow}
\usepackage{multicol}
\usepackage{pdflscape}
\usepackage{afterpage}
\usepackage{everypage}
\usepackage{float}
\usepackage[numbers]{natbib}
\usepackage{algorithm}
\usepackage{algpseudocode}
\usepackage{tikz}%
\usetikzlibrary{positioning}%
\usepackage[utf8]{inputenc}%
\usepackage{caption}%
\usepackage{xurl}
\usepackage{hyperref}
\usetikzlibrary{shapes.geometric, arrows, positioning}

\tikzstyle{state} = [circle, draw=black, thick, minimum size=1.5cm, text centered, text=black, fill=blue!20]
\tikzstyle{arrow} = [thick, ->, >=stealth]

\usepackage{url} 

\begin{document}

\def\spacingset#1{\renewcommand{\baselinestretch}%
{#1}\small\normalsize} \spacingset{1}


\begin{center}
{\Large { \bf Essential Workers at Risk: An Agent-Based Model (SAFE-ABM) with Bayesian Uncertainty Quantification }}
\end{center}

\vspace{.05in} 

\begin{center}
{\large { Elizabeth B. Amona$^{1,2*}$, Indranil Sahoo$^{{2}}$, Ya Su$^{2}$, Edward L. Boone$^{2}$, Gwendoline Nelis$^{3}$, Ryad Ghanam$^{4}$}} \\

\bigskip
{\normalsize {\it $^{1}$UofL Health Brown Cancer Center, University of Louisville, 529 South Jackson Street, Louisville, KY 40202, USA\\
$^{2}$Department of Statistical Sciences and Operations Research, Virginia Commonwealth University \\
$^{3}$Faculty of Public Health, Catholic University of Louvain, Louvain, Belgium\\ 
$^{4}$ Department of Liberal Arts and Sciences, Virginia Commonwealth
University in Qatar, Doha, Qatar\\ }}

\vspace{.1in}

$^*$Corresponding author; Email: \url{elizabeth.amona@louisville.edu}\\

\end{center}

\vspace{.1in}

\vspace{.0075in}
\baselineskip 18truept

\bigskip
\begin{abstract}
 
\noindent Essential workers face elevated infection risks due to their critical roles during pandemics, and protecting them remains a significant challenge for public health planning. This study develops SAFE-ABM, a simulation-based framework using Agent-Based Modeling (ABM), to evaluate targeted intervention strategies, explicitly capturing structured interactions across families, workplaces, and schools. We simulate key scenarios such as unrestricted movement, school closures, mobility restrictions specific to essential workers, and workforce rotation with and without quarantine, to assess their impact on disease transmission dynamics. To ensure robust uncertainty assessment, we integrate a novel Bayesian Uncertainty Quantification (UQ) framework, systematically capturing variability in transmission rates, recovery times, and mortality estimates. Our comparative analysis demonstrates that workforce rotation substantially reduces transmission among essential workers, even without quarantine, while workforce rotation combined with quarantine enforcement provides additional reduction in cumulative deaths and infection burden. Unlike less targeted interventions, the rotational workforce strategies preserve a larger portion of the susceptible population, resulting in a more controlled and sustainable epidemic trajectory. These findings offer critical insights for optimizing intervention strategies that mitigate disease spread while maintaining essential societal functions.

\end{abstract}

\noindent%

\section{Introduction}\label{intro}

Essential workers are individuals who maintain critical societal services despite widespread restrictions during pandemics. This group includes healthcare practitioners such as doctors, nurses, and hospital cleaning staff, as well as food vendors, delivery personnel, and ride-share drivers. During a pandemic, this group faces significantly elevated infection risks due to continued interactions in high-contact environments \cite{milligan2021impact}. Infectious disease outbreaks frequently prompt policymakers to implement broad population-wide interventions, such as school closures, movement restrictions, and social interaction limitations, aimed at curbing general transmission \cite{chung2023evaluating, el2021impact}. Although previous studies have shown these interventions, including lockdown and travel restrictions, effectively reduce infection surges when enacted early \cite{el2021impact, amona2022incorporating}, they are not beneficial for essential workers. This oversight leaves essential workers without specifically tailored protective measures. Despite their crucial societal role, there remains a significant gap in comprehensive modeling that explicitly assesses essential workers' contributions to disease transmission and evaluates targeted protective interventions \cite{milligan2021impact}. Addressing this gap is essential for ensuring equitable and effective pandemic responses.

This situation raises several critical research questions: Does restricting the general population while allowing only essential workers to operate effectively reduce overall transmission, or does it instead concentrate risk within this crucial workforce? Additionally, how do specific intervention strategies such as school closures or mobility restrictions targeting non-essential populations, alter overall transmission dynamics? Could these policies unintentionally elevate transmission risks within workplaces and families? Clearly understanding these trade-offs is vital for designing interventions that reduce disease spread without disrupting essential services.

One proposed strategy for reducing workplace exposure involves splitting essential workers into rotating subgroups to minimize continuous workplace exposure \citep{milligan2021impact}. While rotation may lower transmission within workplaces, it could inadvertently introduce unintended consequences, such as increased workloads on active workers or disruptions in critical service continuity \citep{van2022disruptions}. For instance, healthcare settings have experienced workforce shortages from infection-driven absenteeism, exacerbating facility overcrowding, elevating transmission risks, and placing additional strain on healthcare systems. These complexities highlight the need for a systematic framework for evaluating intervention trade-offs, ensuring policies effectively balance infection control with operational feasibility \citep{phadke2021analysing, mcneill2022extraordinary}.

To systematically explore these trade-offs, this study employs a Structured Agent-based Framework for Essential workers (SAFE-ABM), a simulation-based platform using Agent-Based Modeling (ABM) \citep{bonabeau2002agent, hinch2021openabm, niazi2011agent}. ABMs enable simulations of structured, localized interactions within workplaces, schools, and families, providing insights into how targeted interventions affect disease transmission at both individual and community levels. By explicitly modeling structured interactions, we evaluate various scenarios, including unrestricted movement, school closures, mobility restrictions tailored for essential workers, and workforce rotation schemes. Unlike traditional compartmental models that rely on aggregated population-level assumptions, ABMs capture detailed individual interactions within structured environments, enabling a granular assessment of intervention effectiveness.

A key innovation in our study is the integration of Bayesian Uncertainty Quantification (UQ) within the ABM. Although UQ has previously been explored in epidemiological modeling contexts, its explicit integration within ABMs, particularly utilizing Bayesian inference to quantify transmission uncertainties, remains underdeveloped. Our novel approach systematically evaluates variability in transmission rates, recovery periods, and mortality, ensuring that small fluctuations in parameter estimates do not disproportionately influence intervention conclusions. By employing Bayesian prior distributions informed by domain knowledge, we probabilistically explore epidemic outcomes, while accounting for stochastic agent interactions and epidemiological uncertainty.

ABMs have proven versatile in infectious disease modeling, enabling flexible exploration of transmission dynamics and intervention effectiveness \cite{kerr2021covasim}. Prominent ABM frameworks, including Covasim \citep{kerr2021covasim}, OpenABM-Covid19 \cite{hinch2021openabm}, and FRED \cite{grefenstette2013fred}, have evaluated the effects of social distancing, vaccination, testing, and contact tracing on disease spread. Specifically, Covasim (COVID-19 Agent-based Simulator) is a flexible ABM explicitly designed for COVID-19, capturing detailed demographic structures, comprehensive transmission networks across households, schools, workplaces, and care facilities, and modeling age-specific disease outcomes. Covasim facilitates extensive evaluations of various policy scenarios including physical distancing measures, contact tracing, quarantine protocols, and vaccination strategies, and has been widely used to inform public health decisions globally. Similarly, OpenABM-Covid19 emphasizes computational performance, efficiently modeling structured interactions across households, schools, workplaces, and community settings. Its scalability enables timely large-scale scenario analyses, supporting rapid policy assessments. FRED (Framework for Reconstructing Epidemiological Dynamics), originally developed for influenza modeling and later adapted for COVID-19, integrates realistic synthetic populations with structured interactions across societal layers. It explicitly assesses the impacts of interventions such as vaccination, isolation, and quarantine at local and national scales, providing insights into epidemic dynamics. ABMs like EPINEST \cite{pinotti2024epinest} also offer specialized frameworks for intervention assessment, while CityCOVID \citep{ozik2021population, hack2020us}, explicitly models urban-scale transmission dynamics, employing spatially-resolved contact networks informed by local demographic and behavioral data. ABMs have been successfully applied beyond COVID-19 to influenza \citep{arduin2017agent}, dengue \citep{miksch2019should, mahmood2020agent, deng2008agent}, and Ebola \citep{chen2021hybrid}, demonstrating their adaptability across epidemiological contexts. These models incorporate age-stratified disease progression, spatial transmission networks, and heterogeneous intervention effects, providing comprehensive insights into both pharmaceutical and non-pharmaceutical control measures \citep{lorig2021agent}. 

Despite these significant advances, most existing ABMs simplify workforce heterogeneity, assume uniform social interactions, or generalize intervention effects, thus limiting their effectiveness in evaluating strategies tailored specifically for essential workers. Our model directly addresses these limitations by explicitly incorporating structured movement patterns, targeted interventions, and workforce splitting. Our framework modifies the underlying model structure at each intervention point, reconfiguring agent interactions and movement dynamics to reflect the real-world implementation of policy changes over time. This dynamic restructuring allows us to capture the effects of interventions as they occur, rather than inferring them solely through parameter adjustments, as is typical in many existing ABMs \cite{dyer2023interventionally}. Additionally, our framework integrates detailed family structures to accurately capture intra-family transmission dynamics which allows for precise assessments of how interventions affect transmission within and across structured social environments. Our explicit evaluation of workforce rotation strategies among essential workers further distinguishes this study from the previous ABMs. Essential workers are split into mutually exclusive groups such that while one group is active, the other remains at home, thus modeling a structure that has been proposed in policy discussions but rarely implemented within simulation frameworks.

By explicitly modeling heterogeneous social interactions and incorporating essential worker risk, workforce rotation, and family-level dynamics, our study significantly advances the realism and applicability of ABMs for pandemic response planning. Integrating Bayesian UQ within this essential-worker-focused framework ensures robust intervention evaluations that rigorously account for both stochastic variability and epidemiological uncertainty. These innovations provide crucial insights for optimizing targeted pandemic response strategies, safeguarding essential worker populations, and maintaining societal resilience. The study also explicitly focuses on public health implications rather than economic considerations.

The remainder of this paper is organized as follows: Section \ref{method} presents our the overall statistical methodology, including an overview of the ABM framework, mathematical formulation of the proposed model and key assumptions, simulation scenarios, model algorithm, and uncertainty quantification and validation methods. Section \ref{Result} discusses our simulation results, and Section \ref{conclude} summarizes key findings and highlights future research directions.

\section{SAFE-ABM Modeling Framework}\label{method}

The agent-based model developed in this study, SAFE-ABM (Structured Agent-Based Framework for Essential Workers), simulates disease transmission within structured social environments, explicitly capturing heterogeneous interactions across families, workplaces, and schools. Agents in SAFE-ABM are assigned demographic and behavioral attributes that determine their interactions, movement patterns, and contact structures. By explicitly incorporating individual-level variability, the model offers a detailed representation of the social interactions underlying infection spread.

\subsection{Description of the Agent-Based Model}\label{env}

\begin{figure}[!ht]%
\centering
\includegraphics[width=\textwidth]{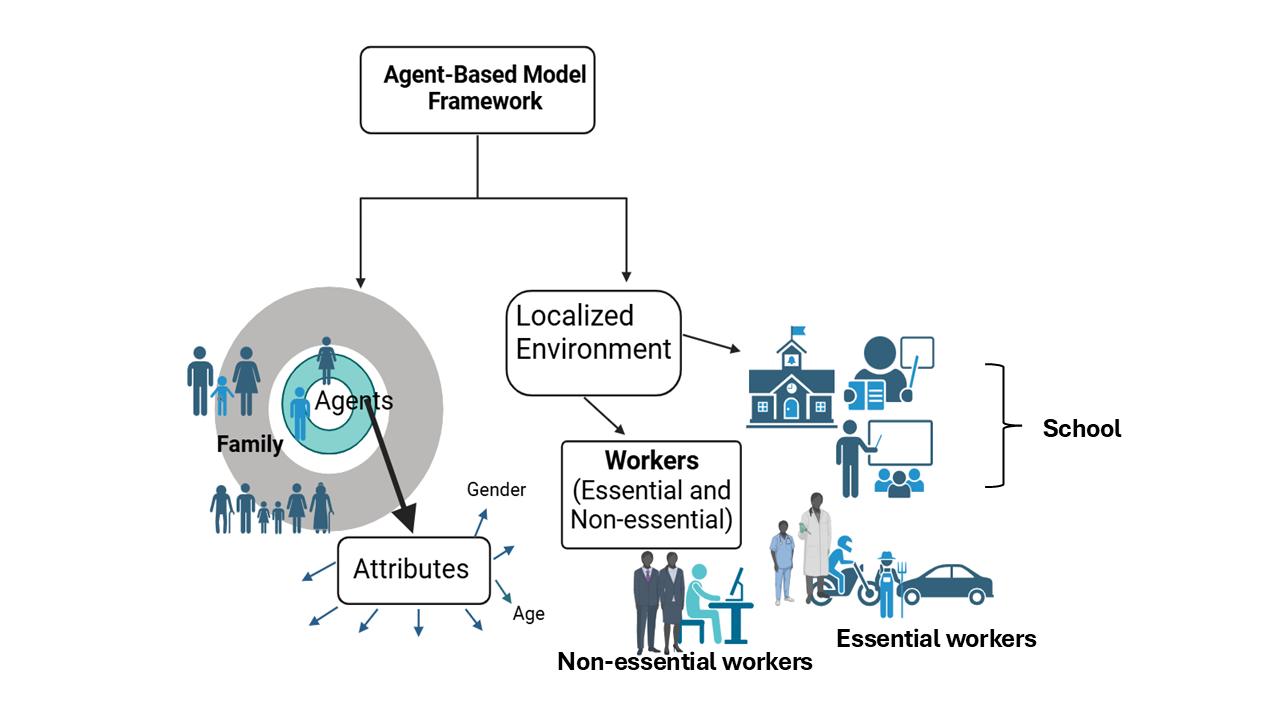}
 \caption{Agent-Based Modeling Framework illustrating agents, attributes, and localized environments such as families, schools (elementary and high school), and workplaces (essential and non-essential workers)}\label{fig:abm_framework}
\end{figure}

Figure \ref{fig:abm_framework} presents the SAFE-ABM framework used in this study to simulate transmission dynamics across structured environments, including families, schools, and workplaces. The model represents a heterogeneous population, where agents interact in localized settings, mimicking real-world contact structures. Each agent is characterized by age, occupation (essential or non-essential worker), and movement behavior, all of which shape their interaction patterns and disease exposure.

At the core of the model is the family unit, where agents are grouped into 2,000 families of varying sizes, forming the primary transmission network. Within families, members engage in repeated, prolonged interactions, making this a high-contact setting for disease spread. Beyond families, schools and workplaces also form structured transmission networks. Schools are stratified into elementary and high school cohorts, facilitating age-dependent mixing patterns. The frequency and intensity of these structured peer interactions vary between elementary and high school students. Workplaces are differentiated by occupation, distinguishing between essential and non-essential workers. Essential workers, including healthcare professionals, delivery personnel, and public service employees, remain active during intervention periods, maintaining workforce participation even under movement restrictions. This sustained occupational exposure contributes to workplace-driven transmission chains. In contrast, non-essential workers experience reduced external exposure, modifying overall disease propagation patterns.

Each agent follows a set of predefined movement and contact rules which are used to calculate the probability of interactions within and across environments. Families represent high-contact networks with frequent interactions, schools facilitate structured peer interactions, and workplaces sustain consistent exposure among colleagues. Essential workers maintain continuous interactions even under restrictive measures, increasing exposure risks in both occupational and household settings. These interactions define the exposure risk landscape within the model, shaping infection pathways across the simulated population. Interventions in the model dynamically modify transmission pathways by adjusting movement dynamics and interaction frequencies. For instance, school closures shift student interactions from schools to families, potentially increasing intra-family transmission. Workforce modifications, such as split essential worker rotations, restructure workplace exposure by alternating work groups, thereby minimizing continuous high-risk interactions. Each intervention scenario captures the trade-offs between mobility restrictions and maintaining essential societal functions, offering insights into how public health measures influence both disease spread and economic stability. 

\subsection{Model Structure and Agent Interactions}
Agents in our model interact dynamically within predefined social structures, with movement patterns and contact networks evolving over time. The first 14 days of the simulation establish baseline interaction dynamics, during which individuals maintain normal mobility and interaction patterns without intervention. During this period, families serve as primary contact units, workplaces allow for occupational interactions, and schools facilitate peer-based mixing. Essential workers remain active throughout the simulation, continuously interacting in both workplace and household environments, while non-essential workers follow routine schedules until movement restrictions are applied.

The implemented interaction rules are summarized in Table~\ref{tab:interaction_rules}. During the first 14 days, all scenarios follow the same baseline contact structure. After day 14, scenario-specific interventions modify the eligible contact environments while preserving the same stochastic disease-transition rules. Under the baseline dynamics, infected agents initiate general random contacts drawn from a Poisson distribution with mean 10, whereas exposed agents initiate general random contacts drawn from a Poisson distribution with mean 15. In addition to these general contacts, transmission can occur through structured household, school, and workplace networks.

\begin{table}[H]
\centering
\caption{Implemented contact structure across intervention scenarios. All scenarios follow the baseline contact structure during days 1--14, while scenario-specific restrictions are applied after day 14.}
\renewcommand{\arraystretch}{1.25}
\begin{tabular}{p{0.24\textwidth} p{0.32\textwidth} p{0.34\textwidth}}
\toprule
\textbf{Contact component} & \textbf{Baseline rule (days 1--14)} & \textbf{Scenario-specific rule (after day 14)} \\
\midrule
General random contacts & Infected agents initiate random contacts drawn from Poisson(10), while exposed agents initiate random contacts drawn from Poisson(15). & Maintained in Scenario 1. In intervention scenarios, random contacts are restricted to agents who remain eligible for external interaction under the scenario rules. \\[0.6em]

Household contacts & Family members are eligible contacts within each household. & Maintained in all scenarios because household transmission remains possible. \\[0.6em]

Elementary school contacts & Elementary students interact within assigned elementary school networks. & Maintained in Scenario 1 and removed in Scenarios 2--5 after day 14. \\[0.6em]

High school contacts & High school students interact within assigned high school networks. & Maintained in Scenario 1 and removed in Scenarios 2--5 after day 14. \\[0.6em]

Non-essential worker contacts & Non-essential workers interact within assigned workplace networks. & Maintained in Scenarios 1 and 2 and restricted in Scenarios 3--5 after day 14. \\[0.6em]

Essential worker contacts & Essential workers interact within assigned workplace networks. & Maintained in Scenarios 1--3. In Scenarios 4 and 5, contacts are restricted to the active rotating essential-worker subgroup.\\[0.6em]

Rotation and quarantine & Not applied during the baseline period. & Rotation is applied in Scenarios 4 and 5, where essential workers rotate on a 7-day schedule. Quarantine of symptomatic active workers is applied only in Scenario 4. \\
\bottomrule
\end{tabular}\label{tab:interaction_rules}
\end{table}

As interventions are introduced, interaction structures adapt accordingly. School closures shift student interactions from classroom settings to family units, altering exposure risks within households. Workforce modifications, such as the split essential worker rotation scheme, reduce occupational exposure by alternating worker schedules, thereby limiting prolonged, high-risk interactions. At each time step, disease transmission is governed by contact-based stochastic processes, where susceptible agents face infection risk based on interaction frequency and the infectious status of their contacts. Upon exposure, individuals enter an incubation period before transitioning to either asymptomatic recovery or symptomatic infection. Symptomatic cases progress to recovery or mortality based on predefined health transition probabilities. The model continuously tracks these transitions, ensuring a dynamic representation of how movement and contact patterns shape epidemic trajectories.

\subsection{Mathematical Framework and Model Assumptions}\label{mathfwk}
\begin{figure}[!ht]
    \begin{center}
        \includegraphics[width=\textwidth]{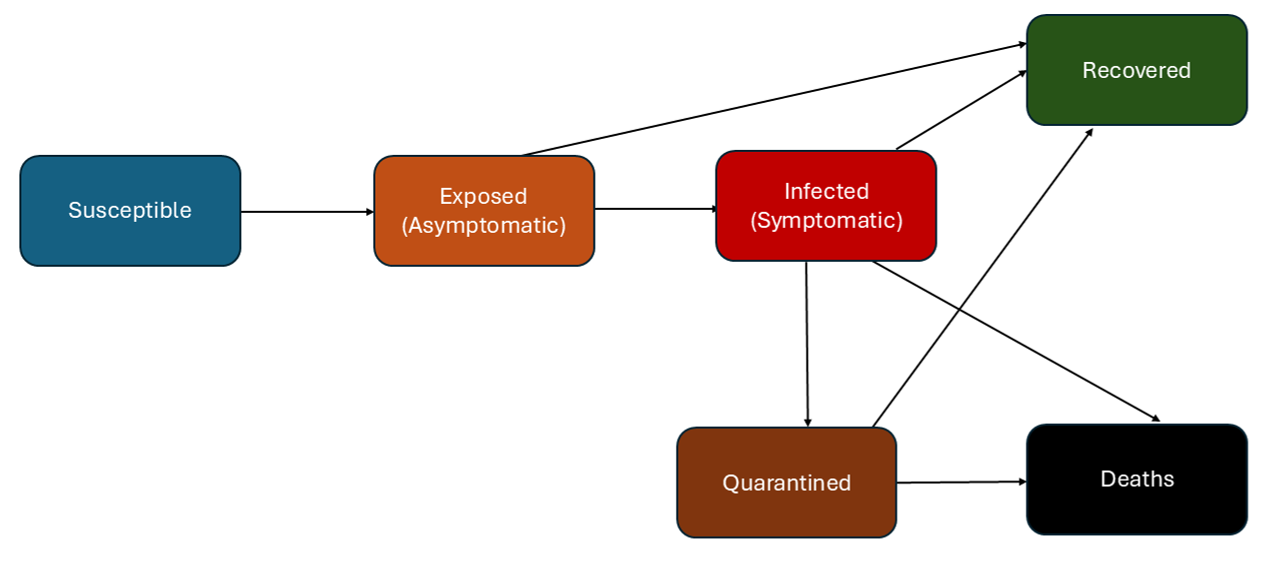}
    \end{center}
    \caption{Epidemiological states and transition pathways in the agent-based model. The model follows a stochastic SEIRD framework, with agents transitioning between states based on probabilistic rules.}
    \label{fig:disease_progression}
\end{figure}

The mathematical framework of SAFE-ABM defines the rules governing disease progression and agent movement dynamics within structured environments. Agents transition between epidemiological states through stochastic processes, with infection, recovery, and mortality determined by predefined probability distributions. Movement patterns establish contact structures within families, workplaces, and schools, thereby shaping exposure risks. Figure \ref{fig:disease_progression} illustrates the epidemiological states and transition pathways implemented within SAFE-ABM, structured by a stochastic Susceptible, Exposed, Infected, Recovered, Quarantined, and Death (SEIRQD) framework.

The model assumes a closed population, with the total number of agents remaining unchanged throughout the simulation. Natural births, deaths, emigration, or immigration are not introduced during the simulation period. Agents begin in the susceptible ($S$) state, meaning they have not yet been exposed to the virus. Upon exposure, they transition to the exposed (asymptomatic) ($E$) state, undergoing an incubation period before either recovering without symptoms or developing symptomatic infection. Symptomatic infected ($I$) individuals may recover, succumb to the disease, or, in scenarios incorporating interventions, be quarantined ($Q$) to reduce further transmission. Recovered ($R$) agents are assumed immune, while deceased ($D$) agents remain in the simulation, ensuring the total population count remains constant.

\subsubsection*{Infection Dynamics}

Susceptible ($S$) agents contract the infection through contact-based transmission upon interaction with an infected individual. The probability of infection is defined as:
\begin{equation*}
    P_{\text{infect}} = 1 - e^{-\beta C I}
\end{equation*}
where \( \beta \sim \mathcal{U}(0.05,0.1) \) is the per-contact transmission probability, \( C \sim \text{Poisson}(\lambda_{\text{env}}) \) denotes the daily contact rate within structured environments (schools, workplaces), and \( I \) is the number of infected (symptomatic) contacts in the agent's network. Infection occurs if a random draw \( U \sim \mathcal{U}(0,1) \) is less than \( P_{\text{infect}} \).

Upon infection, agents transition into the Exposed (Asymptomatic) ($E$) state, entering an incubation period given by:
\begin{equation*}
    \tau_{\text{incubation}} \sim \mathcal{U}(2,15) \text{ days}
\end{equation*}

This incubation-period prior was selected to reflect reported incubation periods across acute epidemic infections relevant to respiratory and emerging viral disease modeling. Table~\ref{tab:incubation_ranges} summarizes reported incubation periods for selected infections. The prior is interpreted as a broad uncertainty range rather than a disease-specific average, allowing the model to represent both short-incubation respiratory infections and moderately longer incubation periods observed in emerging viral infections.

\begin{table}[ht]
\centering
\caption{Reported incubation periods for selected acute epidemic infections used to justify the incubation-period prior.}
\label{tab:incubation_ranges}
\renewcommand{\arraystretch}{1.25}
\begin{tabular}{p{0.22\textwidth} p{0.38\textwidth} p{0.30\textwidth}}
\toprule
\textbf{Disease} & \textbf{Reported incubation period} & \textbf{Source} \\
\midrule
Influenza & Usually 1--4 days after exposure & \cite{CDCInfluenzaYellowBook2026} \\[0.6em]

COVID-19 & Average incubation approximately 5--7 days, with a possible range of 1--14 days & \cite{cimolai2021pursuit} \\[0.6em]

MERS-CoV & Symptoms usually develop approximately 5 days after exposure, with a reported range of 2--14 days & \cite{CDCMERSClinicalOverview2024} \\[0.6em]

Ebola virus disease & Symptoms may appear 2--21 days after exposure, with average onset approximately 8--10 days & \cite{CDCEbolaDiseaseBasics2026} \\[0.6em]

Bundibugyo virus disease outbreak & Incubation period reported as 2--21 days & \cite{WHOBundibugyoDON2026}\\
\bottomrule
\end{tabular}
\end{table}

At the end of incubation, exposed agents either recover without symptoms (\(E \to R\)) with probability:
\begin{equation*}
    P_{E \to R} \sim \mathcal{U}(0.02, 0.03)
\end{equation*}
or progress to symptomatic infection (\(E \to I\)) with probability:
\begin{equation*}
    P_{E \to I} \sim \mathcal{U}(0.05, 0.1)
\end{equation*}
Recovered asymptomatic individuals ($R$) do not contribute to further transmission. Symptomatic infected ($I$) agents subsequently face two possible outcomes: recovery ($I \to R$) with probability:
\begin{equation*}
    P_{I \to R} \sim \mathcal{U}(0.03, 0.07)
\end{equation*}
or mortality (\(I \to D\)) with probability:
\begin{equation*}
    P_{I \to D} \sim \mathcal{U}(0.01, 0.02)
\end{equation*}
Recovered symptomatic individuals gain immunity and no longer contribute to transmission. Agents remain in the Exposed state until they reach the minimum required duration, specifically a sampled incubation period. Once this condition is met, they transition based on the defined probabilities. In contrast, Infected agents do not require a minimum infectious duration; instead, they recover or die based solely on daily transition probabilities. Consequently, the transition probabilities do not sum to 1, as transitions from the Exposed state are only considered after the minimum duration, while transitions from the Infected state may occur at any time.

In the split essential worker rotation with quarantine setting, symptomatic infected agents ($I$) in the active rotating subgroup are immediately quarantined ($Q$) to limit transmission within workplaces. The number of quarantined agents follows:
\begin{equation*}
    Q \sim \text{Poisson}(\lambda_{\text{active}})
\end{equation*}
where $\lambda_{\text{active}}$ represents the expected number of actively infectious essential workers. Quarantined individuals either recover or remain isolated according to predefined transition probabilities. In this setting, $\lambda_{\text{active}}$ dynamically changes throughout the simulation based on the current subgroup of essential workers assigned to on-site work during each rotation cycle. At any given time, only one subgroup is active, and therefore $\lambda_{\text{active}}$ reflects the number of symptomatic infections within that group. This dynamic tracking ensures that only those currently participating in workplace interactions contribute to potential quarantine events, thereby capturing the temporally structured nature of rotational workforce strategies.

\subsubsection*{Agent Movement Rules}\label{movement}

In our SAFE-ABM model, agents interact within structured environments, moving among \textit{families, workplaces, and schools}, where their interactions shape their exposure risk. Each agent is assigned attributes based on \textit{age and occupation}, determining their movement patterns and contact frequency within and across these environments.

Agents belong to family units, forming the primary social structure of the model. A total of \(N_f = 2000\) families are generated, with family sizes (\(S_f\)) following a Poisson distribution:

\begin{equation*}
S_f \sim \text{Poisson}(\lambda_f) + 2
\end{equation*}
where $\lambda_f + 2$ defines the base family size, ensuring a minimum household size of two individuals. Families include adults and children, stratified by age to reflect realistic household compositions. Within families, agents engage in frequent and prolonged interactions, making this setting a primary transmission environment.

Beyond family interactions, workplaces and schools structure additional transmission pathways. A total of \(N_c = 100\) workplaces accommodate adult agents, who are randomly assigned to companies. Interaction frequency within workplaces varies by occupation, with essential workers (including healthcare professionals, delivery personnel, and public service employees) maintaining continuous interactions and higher occupational exposure.

School-aged children are assigned to either elementary or high school environments, with interactions structured by education level. The model includes \(N_e = 20\) elementary schools and \(N_{hs} = 50\) high schools, reflecting realistic age-specific mixing patterns. Elementary students experience higher contact rates, whereas high school students have fewer interactions.

At each time step, corresponding to one day in the simulation, agents move between their assigned family, workplace, or school environments, accumulating exposure based on their contact frequency. The daily number of contacts per agent in each environment follows a truncated Poisson distribution, upper-truncated at the number of other agents, $k_{env}$ present in that environment.
\begin{equation*}
C_{\text{env}}| C_{\text{env}} \leq k_{\text{env}} \sim \text{Poisson}_T(\lambda_{\text{env}})
\end{equation*}
where $\text{Poisson}_T$ denotes a truncated Poisson distribution and $\lambda_{\text{env}}$ is the expected number of daily interactions within a given environment ($\text{env} = \lbrace \text{family}, \text{workplace}, \text{or school} \rbrace$).

\section{Simulation Scenarios}\label{simsc}

To evaluate how movement restrictions and workplace policies influence disease transmission, we simulate multiple scenarios representing varying intervention intensities. These scenarios structure agent interactions across environments, altering exposure risks and transmission pathways. The \textit{No Restriction scenario} serves as the baseline, reflecting unrestricted movement, where agents interact freely without intervention. We refer to the no-restriction baseline as Scenario 1, the school-from-home intervention as Scenario 2, the essential-workers-only intervention as Scenario 3, the rotational workforce strategy with quarantine as Scenario 4, and the rotational workforce strategy without quarantine as Scenario 5. Subsequent scenarios progressively introduce various levels of mobility restrictions and workplace adjustments, enabling comparative analyses of how public health interventions modify movement patterns and shape epidemic trajectories.

\subsection{No Restriction Scenario}

The No Restriction scenario establishes a baseline condition where individuals move freely across environments without interventions. It reflects typical daily activities in the absence of control measures, allowing agents to interact naturally without mobility constraints. This scenario is considered as a baseline for evaluating the effectiveness of subsequent interventions.

\begin{figure}[ht]%
\begin{center}
\begin{tikzpicture}
    \tikzstyle{essential} = [circle, draw=black, thick, minimum size=1.5cm, text centered, text=black]
    \tikzstyle{nonessential} = [circle, draw=black, thick, minimum size=1.5cm, text centered, text=black]

    \node[essential] (E) {E}; 
    \node[nonessential, above right=2cm and 2cm of E] (NE) {NE}; 
    \node[essential, below right=2cm and 2cm of E] (ChildrenE) {(Children) E}; 
    \node[nonessential, below right=2cm and 2cm of NE] (ChildrenNE) {(Children) NE}; 

    \draw[red, thick, ->] (E) to [out=150,in=210,loop] (E);
    \draw[red, thick, ->] (ChildrenE) to [out=150,in=210,loop] (ChildrenE);
    \draw[red, thick, ->] (NE) to [out=30,in=330,loop] (NE);
    \draw[red, thick, ->] (ChildrenNE) to [out=30,in=330,loop] (ChildrenNE);

    \draw[thick, <->] (E) -- (NE);
    \draw[thick, <->] (E) -- (ChildrenE);
    \draw[thick, <->] (NE) -- (ChildrenNE);
    \draw[thick, <->] (ChildrenE) -- (ChildrenNE);
     \draw[thick, <->] (E) -- (ChildrenNE);
\end{tikzpicture}
\caption{Transmission networks for Scenario 1 (No Restriction). Essential workers (E), non-essential workers (NE), and their children interact freely, reflecting an unrestricted movement scenario.}
\label{fig:scenario1}
\end{center}
\end{figure}
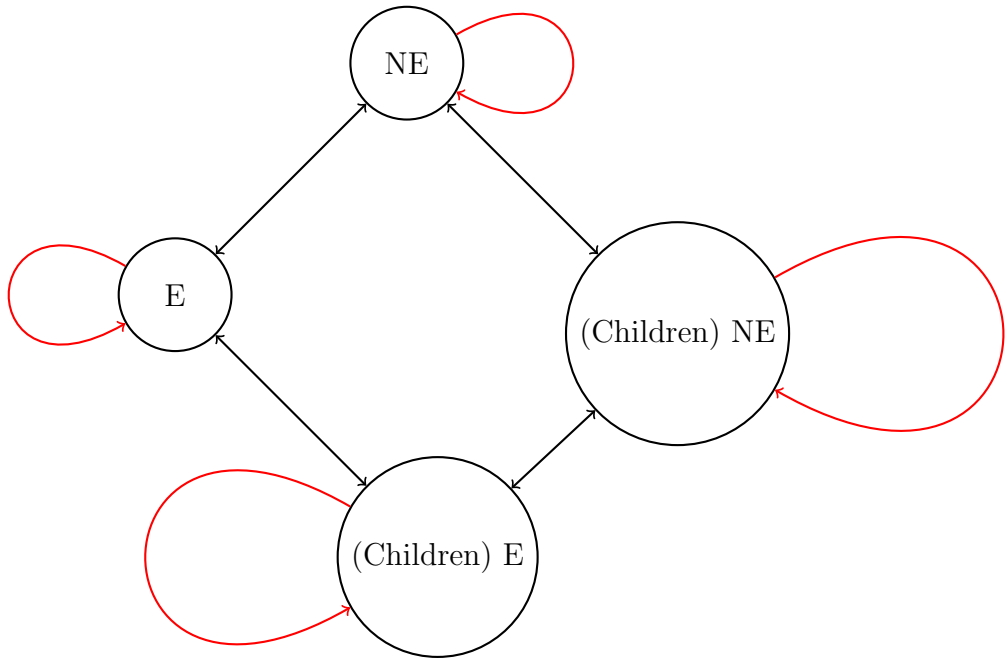

As illustrated in Figure~\ref{fig:scenario1}, agents are categorized as essential (E) and non-essential (NE) workers, and their children are integrated into interaction patterns that mirror real-world social mixing. Essential workers regularly interact within professional environments, primarily engaging with colleagues but also interacting with non-essential workers. This setup represents workplace settings where individuals frequently share spaces and communal facilities. Non-essential workers maintain occupational interactions but typically experience greater flexibility and variability in their external contacts. Children's interaction patterns depend on parental workforce classification. They attend either elementary or high school, following established age-stratified mixing patterns. Schools thus serve as significant transmission hubs due to unrestricted peer interactions contributing to higher overall contact frequency. Additionally, children from essential and non-essential worker families interact freely, simulating natural social behaviors common in educational and recreational settings. The interaction patterns include intra-group contacts (depicted by red self-loops in Figure~\ref{fig:scenario1}), highlighting that agents primarily engage with others within their own category. However, given the absence of movement restrictions, interactions across groups also remain unrestricted. This baseline scenario quantifies the natural progression of disease spread prior to applying interventions and provides a comparative foundation for evaluating subsequent movement restrictions and targeted workplace policies.


\subsection{School from Home Intervention}
We now introduce a targeted intervention in our model under which children remain at home for online schooling, thereby reducing their external contacts. This School-from-Home intervention models a widely implemented pandemic measure intended to minimize children’s exposure by limiting their physical presence in schools and communal spaces.

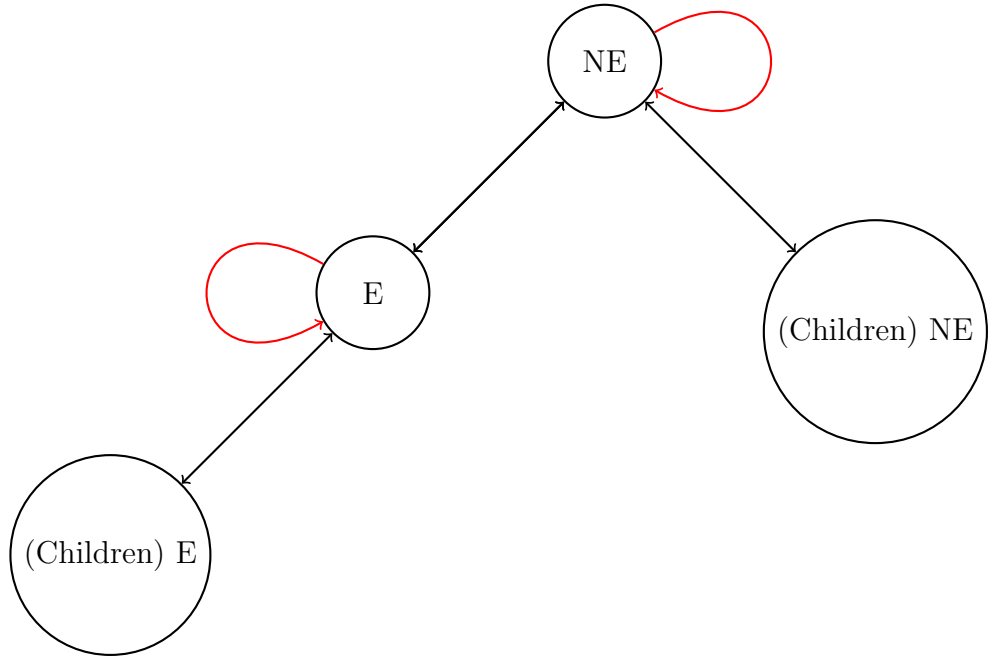
\begin{figure}[ht]
\begin{center}
\begin{tikzpicture}
    \tikzstyle{essential} = [circle, draw=black, thick, minimum size=1.5cm, text centered, text=black]
    \tikzstyle{nonessential} = [circle, draw=black, thick, minimum size=1.5cm, text centered, text=black]

    \node[essential] (E) {E}; 
    \node[nonessential, above right=2cm and 2cm of E] (NE) {NE}; 
    \node[essential, below left=2cm and 2cm of E] (ChildrenE) {(Children) E}; 
    \node[nonessential, below right=2cm and 2cm of NE] (ChildrenNE) {(Children) NE}; 

    \draw[red, thick, ->] (E) to [out=150,in=210,loop] (E);
    \draw[red, thick, ->] (NE) to [out=30,in=330,loop] (NE);

    \draw[thick, <->] (E) -- (NE);       
    \draw[thick, <->] (NE) -- (E);       
    \draw[thick, <->] (E) -- (ChildrenE);   
    \draw[thick, <->] (NE) -- (ChildrenNE);  
\end{tikzpicture}
\caption{Transmission networks for Scenario 2 (School from Home). Essential workers (E), non-essential workers (NE), and their children interact as shown. Children are confined to family units and do not interact across groups, while adults maintain limited interactions within and between groups.}
\label{fig:scenario2}
\end{center}
\end{figure}

In this scenario (Figure~\ref{fig:scenario2}), the population continues to consist of essential workers (E) and non-essential workers (NE), along with their children. However, unlike the No Restriction scenario, children no longer participate in peer interactions, substantially reducing school-based transmission. Instead, their contacts become restricted primarily to immediate family members, removing the cross-group interactions commonly observed within classrooms. The interaction dynamics depicted in Figure~\ref{fig:scenario2} clearly illustrate these structural adjustments. Essential workers maintain their occupational contacts, primarily interacting with other essential workers, as indicated by the self-loops. Similarly, non-essential workers continue interacting predominantly within their own group. Cross-group interactions between essential and non-essential workers persist through shared environments such as workplaces and community settings, highlighting the ongoing role of occupational interactions as potential channels for family-level transmission.

Although children remain isolated from their peers, secondary transmission via parental exposure remains possible. This aligns closely with real-world pandemic scenarios, where children indirectly acquire infections when parents interact externally and subsequently introduce pathogens into family settings. By eliminating direct interactions among children, this intervention strategically disrupts a significant route of virus spread, reinforcing public health recommendations aimed at minimizing transmission risks among school-aged populations \cite{vaivada2022interventions, cdc_school_guidance, cdc_school_health}. The School-from-Home scenario enables the model to assess trade-offs between protecting children from external exposure and maintaining necessary workforce interactions. Despite restricted mobility for children, essential and non-essential workers continue their external activities, emphasizing the role occupational interactions play in shaping family-level transmission patterns.

In parallel, this approach resembles a natural protective mechanism observed among eusocial species. In bee colonies, younger bees typically remain protected within the hive, while older foragers leave to collect resources, thereby reducing exposure risks for vulnerable colony members. Analogously, keeping children home during a pandemic provides protective isolation while still permitting essential workforce participation. This structured strategy effectively balances disease mitigation and societal functionality, mirroring adaptive division-of-labor behaviors documented in nature \cite{seeley1982adaptive, wilson2009genetic, johnson2010division, naug2008structure}.

\subsection{Essential Workers Only Intervention}
Here, only essential workers continue working outside home, while all other individuals remain at home with mobility restrictions. This scenario mimics real-world pandemic policies, where critical services such as healthcare, emergency response, and food distribution must remain operational despite widespread lockdown measures. Essential workers include healthcare professionals, first responders, delivery personnel, and public utility employees, who sustain critical societal functions during health crises \cite{cdc_essential2020, bielicki2020monitoring}.

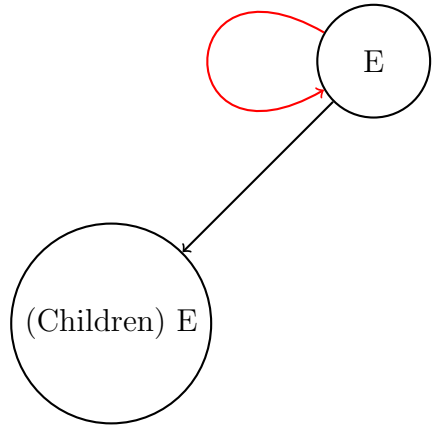
\begin{figure}[!t]
\begin{center}
\begin{tikzpicture}
    \tikzstyle{essential} = [circle, draw=black, thick, minimum size=1.5cm, text centered, text=black]

    \node[essential] (E) {E}; 
    \node[essential, below left=2cm and 2cm of E] (ChildrenE) {(Children) E}; 

    \draw[red, thick, ->] (E) to [out=150,in=210,loop] (E);

    \draw[thick, ->] (E) -- (ChildrenE);   
\end{tikzpicture}
\caption{Transmission networks for Scenario 3 (Only Essential Workers' Movement). Essential workers (E) interact within their group, as shown by the self-loop, and maintain connections with their children. Children are isolated within their family units, with no direct interactions outside these units.}
\label{fig:scenario3}
\end{center}
\end{figure}

Essential workers face increased exposure risks due to persistent workplace interactions. As shown in Figure~\ref{fig:scenario3}, these workers primarily engage in occupational settings, represented by self-loops illustrating frequent and repeated workplace contacts. In environments such as hospitals, transportation networks, and service hubs, regular interpersonal interactions significantly elevate the potential for disease transmission. For example, healthcare workers operating in close proximity may inadvertently spread infections even with protective measures, particularly if safety protocols become compromised \cite{cdc_healthcare2020, who_protection2020, khunti2021assessing, sim2020covid, wei2022risk}. Beyond their occupational settings, essential workers also maintain interactions with external contacts, though typically at reduced frequencies. Service-oriented roles such as food delivery and public transportation involve brief but frequent interactions, providing additional pathways for potential transmission. Although precautionary measures (e.g., contactless delivery) aim to reduce risks, brief customer exchanges still pose a residual risk of disease spread \cite{rahimi2021perceived, nguyen2021effects, romero2021health, karlsson2020covid}. A critical dimension of this scenario involves household transmission risks associated with essential workers. Although children remain at home participating in online schooling, Figure~\ref{fig:scenario3} illustrates how parental occupational exposure can introduce infections into family settings. The directed link between essential workers (E) and their children highlights the elevated secondary transmission risk. Children do not engage in external social activities; however, if parents contract infections through workplace exposure, the family environment becomes a significant route for disease spread \cite{lorenzo2020covid, bi2021insights, karlsson2020covid}.

By isolating non-essential workers and enforcing online schooling, this intervention substantially reduces community-level transmission through limited cross-group interactions. Nonetheless, it emphasizes the occupational burden on essential workers, who face elevated direct exposure risks and an increased likelihood of transmitting infections to their families. Consequently, the scenario underscores the critical importance of workplace protections—including enhanced safety protocols and effective personal protective equipment—to minimize secondary household transmission risks \cite{wang2020household, madewell2020household, wu2020household, madewell2021factors}. Since external mobility is concentrated within this specific group, essential workers represent a key transmission link between occupational and family environments, thereby highlighting workplace safety measures as central components for reducing both direct and indirect disease spread.
\subsection{Split Essential Workers (Rotational Workforce With and Without Quarantine)}

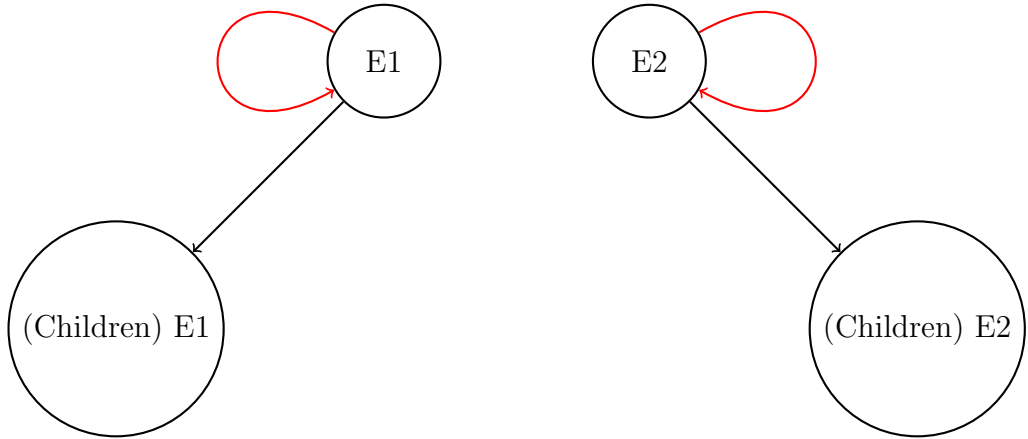
\begin{figure}[ht]
\begin{center}
\begin{tikzpicture}
    \tikzstyle{essential} = [circle, draw=black, thick, minimum size=1.5cm, text centered, text=black]

    \node[essential] (E1) {E1}; 
    \node[essential, below left=2cm and 2cm of E1] (ChildrenE1) {(Children) E1}; 

    \node[essential, right=2cm of E1] (E2) {E2}; 
    \node[essential, below right=2cm and 2cm of E2] (ChildrenE2) {(Children) E2}; 

    \draw[red, thick, ->] (E1) to [out=150,in=210,loop] (E1);
    \draw[red, thick, ->] (E2) to [out=30,in=330,loop] (E2);

    \draw[thick, ->] (E1) -- (ChildrenE1);   
    \draw[thick, ->] (E2) -- (ChildrenE2);   
\end{tikzpicture}
\caption{Transmission networks for Scenarios 4 and 5 (Split Essential Workers: Rotation Scheme). Essential workers are divided into two subgroups (E1 and E2) to alternate their movement and duties. Within each subgroup, essential workers interact among themselves (depicted by self-loops), and they also interact with their children. Scenario 4 applies this rotational structure with prompt quarantine of symptomatic active workers, whereas Scenario 5 applies the same rotational structure without quarantine}. This setup minimizes sustained high-risk interactions across the entire essential workforce while maintaining household exposure dynamics. \label{fig:scenario4}
\end{center}
\end{figure}

The final intervention category, represented by Scenarios 4 and 5 (Figure~\ref{fig:scenario4}), implements a rotation system where essential workers are divided into two mutually exclusive subgroups, E1 and E2, alternating weekly between active duty and staying home. These scenarios reduce workplace density while maintaining essential services during a pandemic. Each subgroup operates on a 7-day alternating schedule: one subgroup (say, E1) actively works for one week while the other subgroup (E2) remains at home, with no interaction between groups during this period. In Scenario 4, if workers in the actively working subgroup become symptomatic, they are promptly quarantined to prevent interaction with healthy workers, further reducing transmission risks. Scenario 5 applies the same rotational structure without quarantine. After seven days, roles reverse; E2 becomes active, and E1 stays home. This structured, mutually exclusive rotation explicitly tests whether reducing the number of concurrently active workers, with or without rapid quarantine of symptomatic individuals, effectively reduces disease transmission, or if it unintentionally increases risks due to concentrated interactions within smaller groups.

As illustrated in Figure~\ref{fig:scenario4}, when subgroup E1 is active, members interact among themselves, depicted by the self-loop around E1, and subsequently return home to their families (indicated by the arrow from E1 to `Children E1'). During the same period, the inactive subgroup (E2) stays home, interacting solely with household members. After one week, the subgroups switch roles, and this alternating pattern continues throughout the simulation. 
Although rotation schemes are intended to lower workplace density, they can inadvertently increase interaction frequency within the actively working subgroup. Unlike situations where the entire workforce is active simultaneously, splitting the workers results in more frequent interactions among a smaller group of colleagues during each rotation. Real-world studies indicate that workforce rotations might unintentionally amplify infection rates if interactions among active subgroup members intensify \cite{sah2018disease, sim2020covid}.

We hypothesize that while rotational workforce strategies effectively reduce overall workplace exposure, increased interactions within active subgroups could raise within-group transmission risks. However, we also hypothesize that enforcing quarantine for symptomatic individuals could effectively mitigate these heightened risks, resulting in an overall reduction in workplace and secondary household transmission.

Given that rotation alone might inadequately control transmission risks and could potentially intensify exposure among active workers, we incorporate an additional measure: quarantining symptomatic workers. Quarantine ensures symptomatic workers are quickly removed from circulation, limiting further disease spread within the subgroup. By promptly isolating infected individuals before transmission can occur extensively, quarantine effectively interrupts transmission chains. This practice aligns closely with historical and contemporary public health strategies, where quarantine remains fundamental for controlling infectious disease spread \cite{tognotti2013lessons}. Combining workforce rotations with targeted quarantine thus allows evaluation of whether a hybrid intervention is more effective at mitigating workplace and household transmission compared to strategies that solely involve essential workers working continuously.

This intervention introduces two significant changes in exposure dynamics. First, dividing essential workers into rotating subgroups might increase intra-group transmission risks due to more frequent interactions among a smaller, actively working group. Because only half of the essential workforce is active at any given time, interactions within each subgroup become more frequent, potentially intensifying transmission risks. Second, introducing quarantine as a complementary measure provides a counterbalance, promptly removing symptomatic workers and preventing subgroup-level outbreaks. The effectiveness of rotation in reducing overall exposure therefore heavily depends on how efficiently quarantine measures limit transmission among active subgroup members.

Contrasting Scenario 3, where all essential workers remain continuously active, Scenario 4 attempts to disrupt prolonged workplace transmission by implementing alternating work schedules combined with quarantine measures, while Scenario 5 evaluates the same rotational structure without quarantine. By explicitly modeling these competing dynamics, Scenarios 4 and 5 allow us to evaluate whether rotating essential workers, with or without quarantine, can reduce both workplace outbreaks and secondary household infections. Results from these scenarios can directly inform public health policies designed to optimize workforce safety while maintaining essential societal functions during pandemics. By assessing structured workforce rotations with and without quarantine practices, the model provides key insights into effectively managing workplace outbreaks, limiting secondary family transmission, and balancing infection control with continued economic and societal stability. These insights emphasize the value of targeted public health measures aimed specifically at protecting high-risk workers while ensuring critical infrastructure remains operational, thereby providing valuable guidance for developing effective pandemic response frameworks.

\subsection{Algorithm}\label{algo}

Algorithm~\ref{algoo} outlines our SAFE-ABM model by explicitly structuring agent movements, infection dynamics, and intervention mechanisms within a computational framework. The algorithm implementation aligns precisely with the mathematical model described in Section~\ref{mathfwk} and incorporates the distinct intervention strategies detailed in Subsections~3.1--3.16. By systematically integrating these components, the algorithm ensures that each simulated scenario accurately represents the dynamic effects of policy changes on disease transmission within heterogeneous populations. Specifically, it captures the dynamic restructuring of agent interactions under various intervention conditions, enabling clear comparisons of epidemic outcomes across scenarios.

\begin{algorithm}[H]
\captionsetup{font=small}
\caption{Agent-Based Model (ABM) for Epidemic Spread}\label{algoo}
\begin{algorithmic}[1]
\State \textbf{Step 1: Initialization}
 \State \textit{Input:} Population size $N = 10,000$, Families $n_{\text{fam}} = 2000$
 \State \textit{Output:} Agent population with attributes
 \State \textbf{Assign agent attributes}:
\noindent $\rightarrow$ Unique ID, Age, Gender (\{M, F\}), Cohort (\{E, NE, HS, e\}) \\
\noindent $\rightarrow$ NE = Non-Essential Worker, E = Essential Worker, HS = High School, e = Elementary Student
\State Assign Family, School, and Work Networks       
\State Assign meeting frequencies for family, work, and school
\State Save the initialized population

\State \textbf{Step 2: Disease Seeding}
\State \textit{Input:} Initialized population
\State \textit{Output:} Initial infected and exposed individuals
\noindent $\rightarrow$ Exposed: $E_0 = 10$ agents \quad $\rightarrow$ Infected: $I_0 = 5$ agents \\
\noindent $\rightarrow$ Compute initial susceptible population: $S_0 \gets N - E_0 - I_0$
\State \textbf{Step 3: Epidemic Dynamics}
\State \textit{Input:} Initial conditions $(S_0, E_0, I_0)$, time steps $t = 1, \dots, T$
\State \textit{Output:} Time-series $(S_t, E_t, I_t, R_t, D_t)$
\State Define transmission parameters:
\noindent $\rightarrow$ Transmission Probability: $\beta \sim \text{Uniform}(0.05, 0.1)$ \\
\noindent $\rightarrow$ Exposure-to-Infection Probability: $P_{E \to I} \sim \text{Uniform}(0.05, 0.1)$ \\
\noindent $\rightarrow$ Exposure-to-Recovery Probability: $P_{E \to R} \sim \text{Uniform}(0.02, 0.03)$ \\
\noindent $\rightarrow$ Recovery Rate: $P_{I \to R} \sim \text{Uniform}(0.03, 0.07)$ \\
\noindent $\rightarrow$ Death Rate: $P_{I \to D} \sim \text{Uniform}(0.01, 0.02)$
\For{$t = 1$ to $T$}
    \State \textbf{Transmission Dynamics:}
    Each infected agent ($I$) contacts a number of others drawn from a truncated Poisson distribution, 
    $C_{\text{env}}| C_{\text{env}} \leq k_{\text{env}} \sim \text{Poisson}_T(\lambda_{\text{env}})$, 
    upper-bounded by the number of other agents, $ k_{\text{env}}$ in the environment.
    \State Infection is governed by the probabilistic function 
    $P_{\text{infect}} = 1 - e^{-\beta C I}$, where $\beta$ is the per-contact transmission probability 
    and $C$ is drawn as described under Infection Dynamics.
    \State Update exposed population:
    \[
    E_{t+1} \gets E_t + \sum_{i \in I} \mathbb{I}(U_i < P_T)
    \]
    \State \textbf{State Transitions:}
    \State Exposed $\to$ Infected with probability $P_E$
    \State Infected $\to$ Recovered with probability $\gamma$
    \State Infected $\to$ Deaths with probability $\mu$
    \State Update susceptible population:
    \[
    S_{t+1} \gets S_t - \big(E_{t+1} - E_t\big)
    \]
\EndFor
\State \textbf{Step 4: Intervention Phase (After $t > 14$)}
\noindent $\rightarrow$ \textbf{No Restriction:} Continue baseline dynamics \\
\noindent $\rightarrow$ \textbf{School-From-Home:} Remove school-based contacts \\
\noindent $\rightarrow$ \textbf{Essential Workers Bubble:} Restrict non-essential workers' contacts \\
\noindent $\rightarrow$ \textbf{Split Essential Workers with Quarantine:} Rotate essential workers in subgroups and quarantine symptomatic active workers\\
\noindent $\rightarrow$ \textbf{Split Essential Workers without Quarantine:} Rotate essential workers in subgroups without applying quarantine
\State \textbf{Return:} Time-series $(S_t, E_t, I_t, R_t, D_t)$
\end{algorithmic}
\end{algorithm}

\section{Uncertainty Quantification and Model Validation}\label{UQMV}

Ensuring the reliability of simulation outcomes requires a rigorous approach to uncertainty quantification, parameter-range selection, and simulation-based validation. This section introduces the Bayesian Uncertainty Quantification (UQ) framework implemented in our SAFE-ABM model to systematically account for parameter variability and stochastic transmission dynamics. We also outline the parameter-range selection process, simulation-based internal validation approach, and sensitivity analysis used to support robust statistical inference in intervention assessments.

\subsection{Bayesian Uncertainty Quantification (UQ) Framework}
A few studies have explored the use of Bayesian methods for uncertainty quantification in agent-based modeling. \cite{liu2022bayesian} developed a Bayesian framework to quantify uncertainty in ABMs of networked group anagram games, which focuses on clustering diverse player behaviors rather than epidemic dynamics. Their approach integrates Bayesian nonparametric clustering and multinomial regression to model behavioral transitions, making it well-suited for decision-making dynamics rather than structured epidemiological processes. \cite{mcculloch2022calibrating} addressed ABM calibration by combining History Matching and Approximate Bayesian Computation to refine parameter spaces and provide credible intervals. While their approach enhances parameter estimation and model fitting, it does not focus on uncertainty propagation in epidemic modeling.

In contrast, our study introduces a Bayesian uncertainty quantification framework specifically tailored to \textit{epidemiological agent-based models}. Unlike clustering-based and calibration-focused approaches, our framework directly propagates uncertainty through stochastic simulations, systematically quantifying intervention impacts on epidemic trajectories. By simulating 100 independent runs with distinct population samples, our framework captures the variability introduced by heterogeneous population structures and stochastic transmission dynamics. This structured approach enables a rigorous assessment of uncertainty in disease outcomes, offering novel insights into intervention effectiveness within agent-based epidemiological models.

Our Bayesian UQ framework assigns prior distributions to key epidemiological parameters \cite{martinez2014trends, maclehose2014applications}, explicitly capturing variability in disease transmission, recovery, and mortality rates, thus yielding probabilistic epidemic trajectories rather than deterministic outcomes. To reflect plausible disease dynamics, we define prior distributions over core transmission parameters:

\begin{align}\label{eq:1}
\text{Transmission Probability} &\sim U(0.05, 0.1) \\ \nonumber
\text{Recovery Rate} &\sim U(0.03, 0.07) \\ \nonumber
\text{Mortality Rate} &\sim U(0.01, 0.02) 
\end{align}
 For each simulation run, values of the epidemiological parameters in Eq.~\ref{eq:1} were sampled from their specified uniform prior distributions, with the same prior ranges applied across all intervention scenarios to ensure comparable scenario evaluation.

Each simulation explicitly captures stochastic interactions among agents (intra-run variability) and differences in epidemiological parameters and population structures across independent runs (inter-run variability), providing a comprehensive uncertainty assessment. This methodology enhances model robustness by incorporating prior knowledge and uncertainty directly into the agent-based framework.

To fully capture uncertainty, we conduct 100 independent simulation runs, each initialized with a unique synthetic population realization. At initialization, 15 agents were randomly selected from the full synthetic population, with 10 assigned to the exposed state and 5 assigned to the infected state, using the same outbreak-seeding procedure across all scenarios. No additional outbreak takeoff criterion or reseeding procedure was imposed. This design ensures that observed variability in outcomes is not solely due to stochastic agent interactions but also accounts for differences in underlying population structures and parameter variability. The Force of Infection (FoI), along with state variables (Susceptible, Exposed, Infected, Recovered, Deaths), is recorded across runs, facilitating post-simulation statistical analysis of uncertainty bounds.

To efficiently handle large-scale uncertainty propagation, the simulation workflow was optimized and parallelized using Python's \texttt{ProcessPoolExecutor} from the \texttt{concurrent.futures} module. For each intervention scenario, we generated 100 distinct synthetic populations, each consisting of 10,000 agents and simulated over 120 days. This design allowed uncertainty to reflect both sampled epidemiological parameters and variation across population realizations. The optimized implementation was executed on a multi-core workstation using 37 processing cores. The no-restriction scenario required the longest computational time, approximately 8 hours, while the school-from-home scenario required approximately 5 hours; the remaining optimized scenarios completed more quickly under the same parallel-processing setup.

The computational implementation of our Bayesian uncertainty quantification framework is summarized in Algorithm \ref{algoUQ}, which outlines the structured approach used to quantify parameter variability and stochastic uncertainty across multiple simulation runs.

\begin{algorithm}[H]%
\captionsetup{font=small}
\caption{Uncertainty Quantification (UQ) in Agent-Based Model (ABM)}\label{algoUQ}
\begin{algorithmic}[1]
\State \textbf{Step 5: Initialization}
    \State \textit{Input:} Number of simulation runs per seed $n_{\text{sim}}$, Number of seeds $n_{\text{seeds}}$, Time steps $T$
    \State \textit{Output:} Quantified uncertainty estimates for epidemic dynamics
    \vspace{0.05cm}
    \State Define Bayesian priors for epidemic parameters:
    \begin{itemize}
        \item[$\rightarrow$] Per-contact transmission probability: $\beta \sim \mathcal{U}(0.05, 0.1)$
        \item[$\rightarrow$] Recovery rate: $\gamma \sim \mathcal{U}(0.03, 0.07)$
        \item[$\rightarrow$] Death rate: $\mu \sim \mathcal{U}(0.01, 0.02)$
    \end{itemize}

    \State Sample distinct agent populations for each simulation run ($N = 10,000$ per run)
    \vspace{0.05cm}
    
\State \textbf{Step 6: Scenario-Based Simulation}
    \State \textit{Input:} Initialized agent populations, priors
    \State \textit{Output:} Time-series epidemic data for all scenarios
    \For{$\text{seed} = 1$ to $n_{\text{seeds}}$}
        \For{$\text{sim} = 1$ to $n_{\text{sim}}$}
           \State Sample $(\beta, \gamma, \mu)$ from prior distributions
            \State Load agent population for current seed
            \State \textbf{Run scenario-based simulation:}
            \begin{itemize}
                \item [-] \textbf{No Restriction:} Baseline transmission dynamics
                \item [-] \textbf{School From Home:} Remove school-based contacts
                \item [-] \textbf{Essential Workers Only:} Restrict non-essential workers' interactions
                \item [-] \textbf{Split Essential Workers with Quarantine:} Alternate essential workers in rotations of 7-day shifts and quarantine symptomatic active workers
                \item [-] \textbf{Split Essential Workers without Quarantine:} Alternate essential workers in rotations of 7-day shifts without applying quarantine
            \end{itemize}
            \State Record epidemic trajectories $(S_t, E_t, I_t, R_t, D_t)$ for each run
        \EndFor
    \EndFor
    \vspace{0.05cm}
    
\State \textbf{Step 7: Aggregation of Results}
    \State \textit{Input:} Time-series data from multiple runs
    \State \textit{Output:} Uncertainty estimates (quantiles)
    \State Estimate quantiles $(q_{2.5}, q_{50}, q_{97.5})$ for:
    \begin{itemize}
        \item [$\rightarrow$] Exposed $(E_t)$
        \item [$\rightarrow$] Infected $(I_t)$
        \item [$\rightarrow$] Recovered $(R_t)$
        \item [$\rightarrow$] Deaths $(D_t)$
        \item [$\rightarrow$] Force of Infection
    \end{itemize}
    
\State \textbf{Step 8: Save and Analyze Results}
    \State Store quantified uncertainty estimates for further Bayesian inference
    \State \textbf{Return:} Uncertainty estimates for epidemic states
\end{algorithmic}
\end{algorithm}

\subsection{Parameter-Range Selection and Plausibility Assessment}

To ensure the robustness of our SAFE-ABM model and the reliability of simulated outcomes, we implement parameter-range selection, simulation-based validation, and sensitivity analysis within a Bayesian uncertainty quantification (UQ) framework. Since our study uses synthetic populations rather than empirical epidemiological data, a structured simulation-based validation approach is crucial to maintain statistical rigor and avoid biases from single-run stochastic variability. The simulation framework, including the generation of synthetic populations and model initialization, was implemented in Python version 3.12.7~\cite{python312}.

This study did not perform formal empirical calibration to an external ground-truth epidemic dataset. Instead, parameter ranges were selected through exploratory simulation-based plausibility assessment. Candidate ranges were evaluated by repeated simulations to determine whether they produced qualitatively reasonable epidemic dynamics, including an initial growth phase, a discernible epidemic peak, and a subsequent decline. Parameter ranges that produced immediate extinction, highly unstable trajectories, or unrealistically rapid saturation across most simulations were not retained. Thus, the term ``plausible'' refers to qualitative consistency with expected epidemic progression rather than agreement with a specific observed outbreak dataset.

To ensure fair comparisons across intervention strategies, all scenarios employ identical prior distributions for the epidemiological parameters. Parameters are drawn from uniform distributions within the retained ranges so that differences in outcomes across scenarios reflect differences in intervention structure under the same epidemiological assumptions, rather than scenario-specific parameter choices.

Model validation in this study refers to simulation-based internal validation rather than empirical validation against an external observed outbreak dataset. Specifically, we assessed whether the SAFE-ABM framework produced coherent and consistent system-level epidemic behavior across repeated stochastic simulations and distinct synthetic populations. For each scenario, 100 independent simulation runs were performed, with each run initialized using a distinct synthetic population and sampled epidemiological parameters.

The validation assessment focused on trajectory-level consistency across susceptible, exposed, infected, recovered, and deceased states. Median trajectories and 2.5\%--97.5\% simulation-based uncertainty intervals were computed to evaluate whether scenario outcomes were stable across repeated runs rather than being driven by a single stochastic realization. Thus, the validation approach was based on internal simulation consistency, uncertainty quantification, and robustness across population seeds, not on calibration or validation against a specific empirical ground-truth dataset.

We analyzed cumulative recoveries and cumulative deaths as downstream resolved-outcome measures of epidemic burden. Although cumulative infection burden is central to transmission dynamics, cumulative recoveries and deaths provide complementary information about how infections resolve over time. Cumulative deaths represent the severe public-health consequence of transmission, while cumulative recoveries reflect the number of agents who passed through infection and survived. Because recovery and death probabilities were drawn from the same prior ranges across scenarios, differences in these outcomes arise from scenario-specific transmission structure and timing rather than scenario-specific outcome probabilities. We therefore interpret cumulative recoveries and deaths alongside the infected trajectories rather than as independent measures of transmission alone.

Sensitivity analysis further evaluated the stability of model conclusions under parameter variability. Key epidemiological parameters--including transmission probability, minimum exposure duration, and recovery rate--were systematically varied within their prior ranges to assess impacts across different scenarios. In addition, the use of distinct synthetic populations across simulations naturally captures heterogeneity in population structures and stochastic transmission pathways. This approach supports robust comparison of intervention strategies across a broad range of plausible simulation conditions. By combining parameter-range selection, simulation-based internal validation, uncertainty quantification, and sensitivity analysis, we strengthen the credibility and reliability of SAFE-ABM for evaluating intervention strategies under uncertainty.

To support numerical comparison across intervention scenarios, we computed inflection points, final cumulative recovery and death summaries, pairwise Welch's t-tests \citep{welch1947generalization}, and mutual information sensitivity rankings \citep{shannon1948mathematical, kraskov2004estimating}. Inflection points were computed numerically from the cumulative recovery and death trajectories as the time point at which the rate of change began to decline. For a cumulative trajectory $f(t)$, the daily increment was defined as $\Delta f(t)=f(t)-f(t-1)$. The inflection day was identified when the daily increment at the next time point became smaller than the current daily increment, that is, when $\Delta f(t+1)<\Delta f(t)$, equivalently $f(t+1)-f(t)<f(t)-f(t-1)$. This finite-difference criterion provides a discrete-time analogue of the second-derivative criterion and directly captures the point at which the rate of change starts to decline. The criterion was applied to the scenario-specific median cumulative trajectories shown in Figures~\ref{fig:recovScenarios} and~\ref{fig:deathsScenarios} to obtain the inflection day reported for each scenario and outcome. The same criterion was applied separately across the 100 run-level trajectories to summarize the mean, standard deviation, and 95\% interval of the inflection days.

Final cumulative recoveries and deaths were computed at the final simulation time point and summarized using the mean and standard deviation across the 100 simulation runs. Pairwise two-sided Welch's t-tests were used to compare scenario-specific outcome distributions, allowing for unequal variances. Statistical significance was assessed at $\alpha=0.05$.

To assess input-output sensitivity, mutual information was computed between each model input and selected simulation outputs using \texttt{mutual\_info\_regression} from scikit-learn \citep{pedregosa2011scikit}. The input features included the sampled epidemiological parameters and the seed index, while the outputs included final cumulative recoveries, final cumulative deaths, recoveries at day 30, and deaths at day 30. Larger mutual information values indicate stronger association between an input feature and variation in the corresponding output.

While our simulations explore plausible parameter ranges and epidemic behaviors through exploratory analyses, formal calibration to empirical data remains outside the scope of this study. This decision reflects our immediate focus: to develop, implement, and validate a flexible agent-based framework capable of realistically simulating structured interventions. Future work will extend this framework by incorporating empirical epidemiological data for full parameter calibration, enabling more precise, data-driven policy assessments.

\section{Results}\label{Result}
\subsection{Epidemic Progression Across all Scenarios}
\begin{figure}[H]
 \begin{subfigure}[t]{0.5\textwidth}
  \centering
     \includegraphics[width=\textwidth]{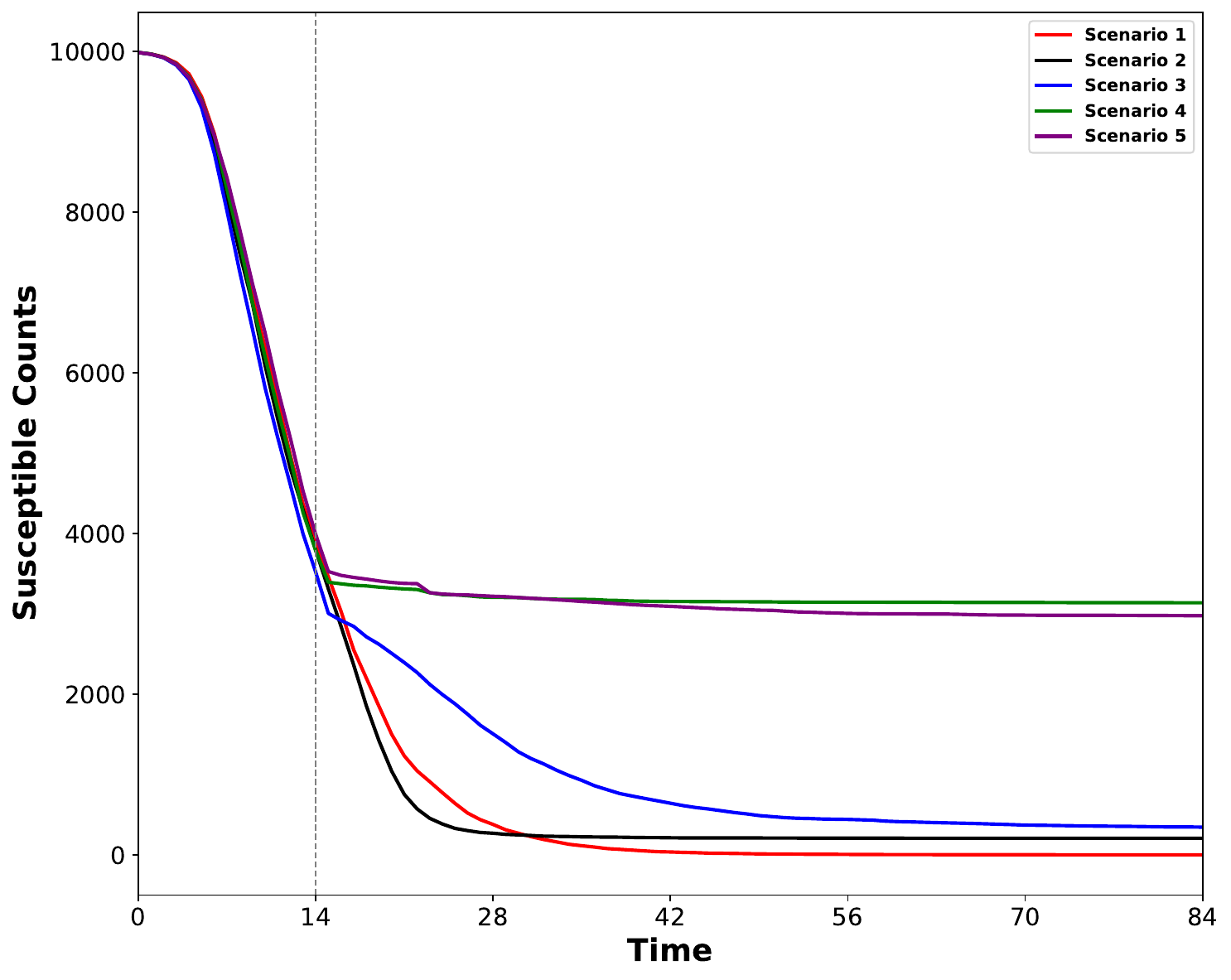}
     \caption{Susceptible Counts}
     \label{fig:sucept}
 \end{subfigure}
 \hfill
 \begin{subfigure}[t]{0.5\textwidth}
  \centering
     \includegraphics[width=\textwidth]{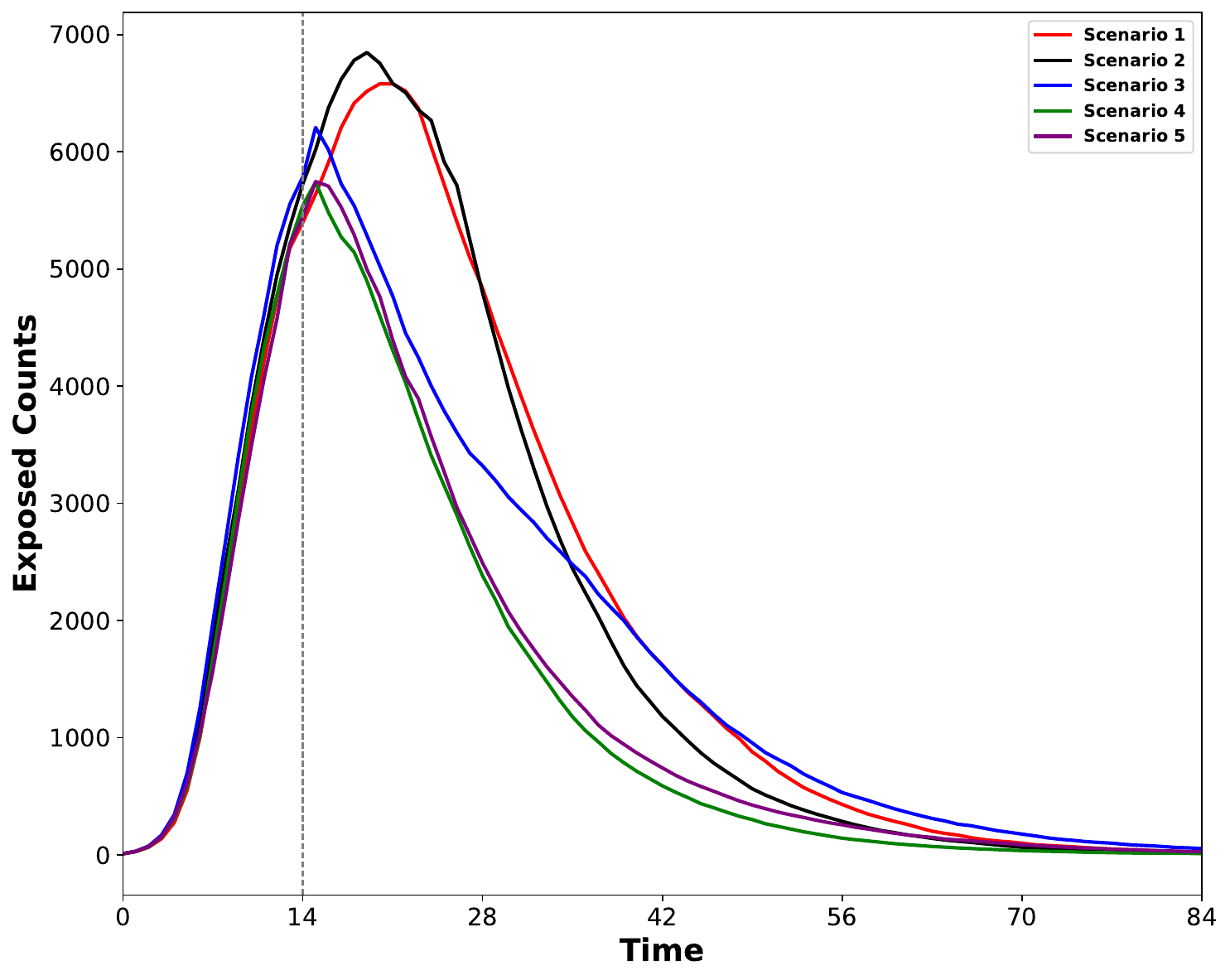}
     \caption{Exposed (Asymptomatic) Counts}
     \label{fig:exp}
 \end{subfigure}
 \begin{subfigure}[t]{0.5\textwidth}
  \centering
    \includegraphics[width=\textwidth]{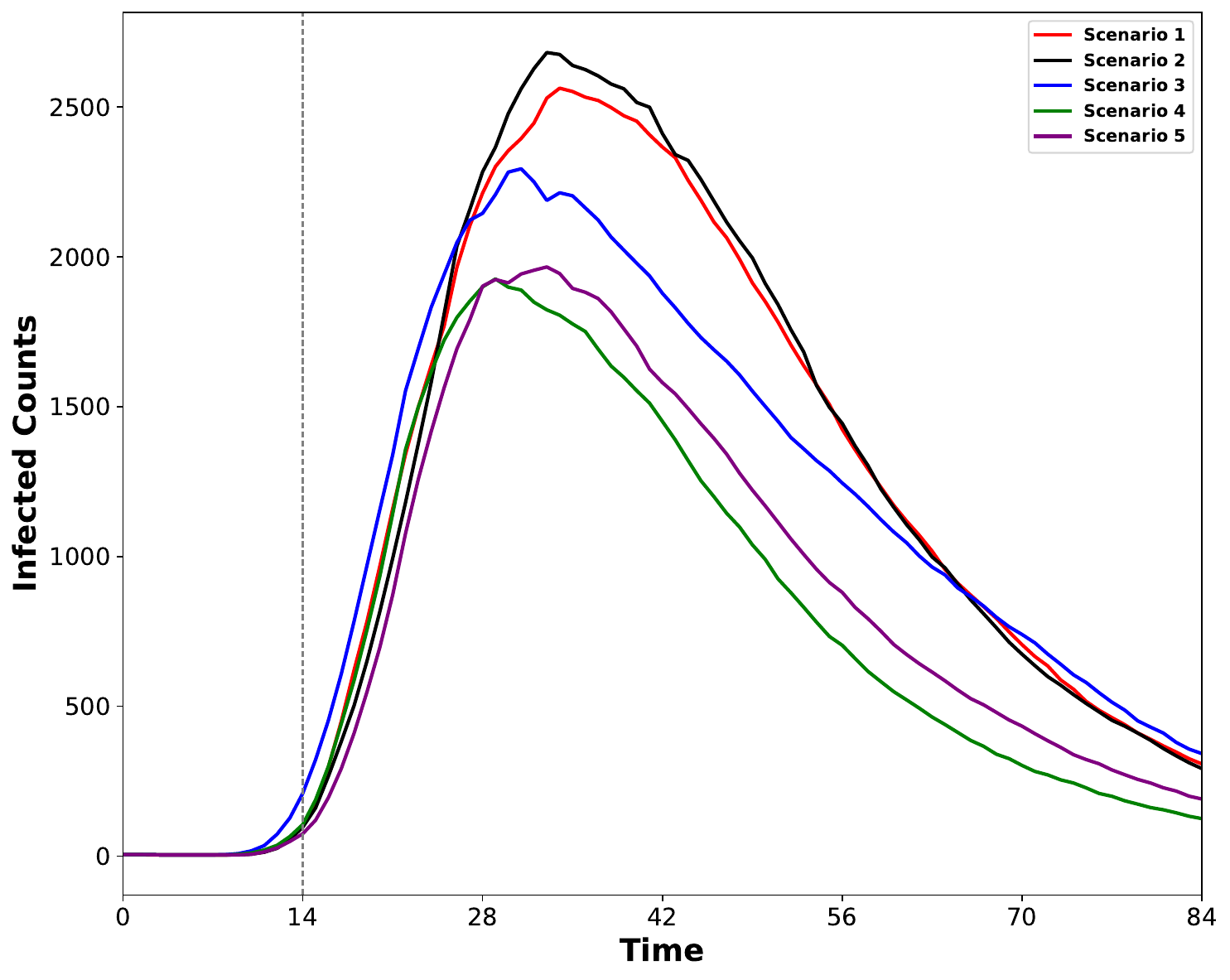}
     \caption{Infected (Symptomatic) Counts}
     \label{fig:inf}
 \end{subfigure}
 \hfill
 \begin{subfigure}[t]{0.5\textwidth}
  \centering
    \includegraphics[width=\textwidth]{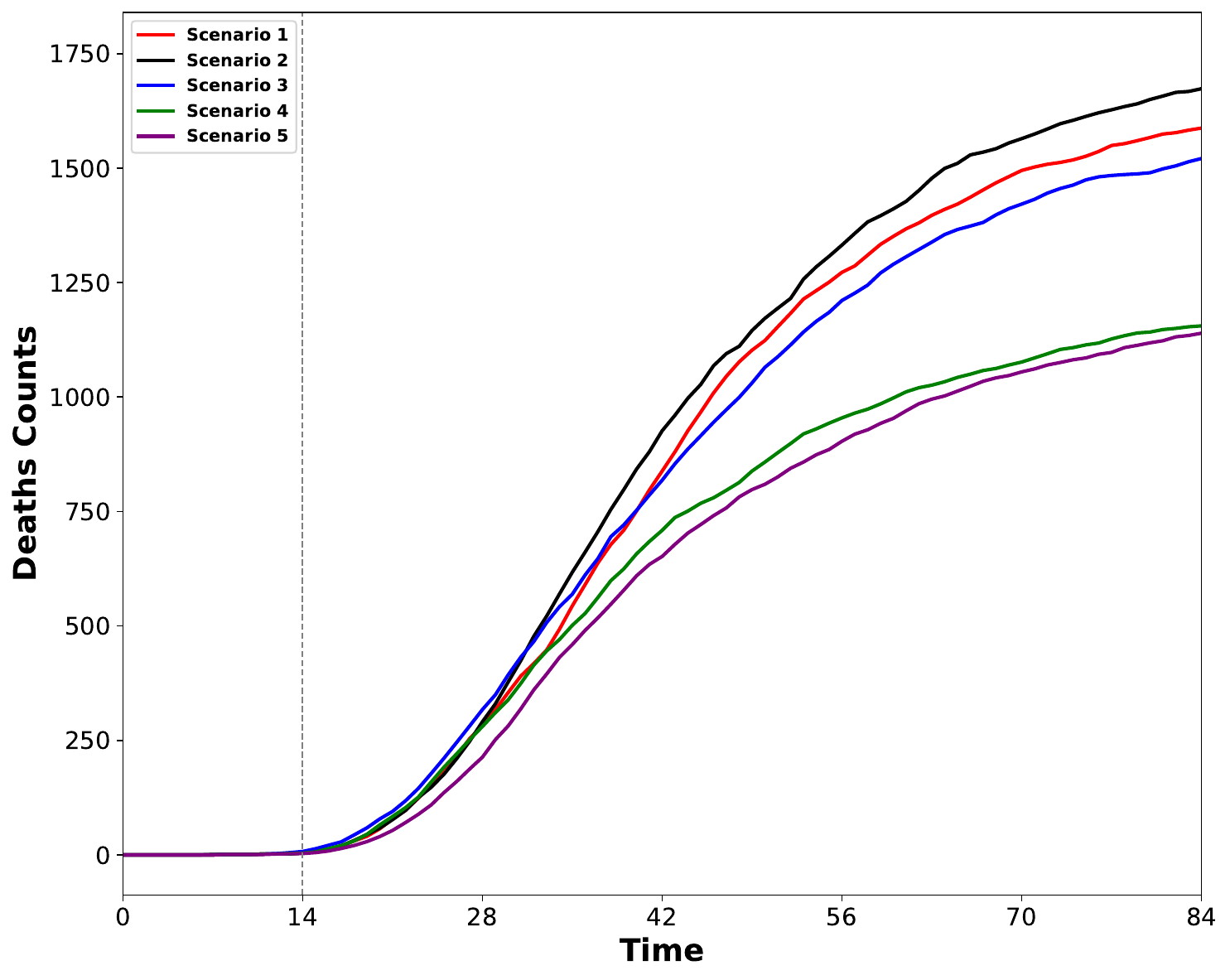}
     \caption{Deaths}
     \label{fig:dea}
 \end{subfigure}
 \hspace*{\fill} 
 \begin{subfigure}[t]{0.5\textwidth} 
  \begin{center}
     \includegraphics[width=\textwidth]{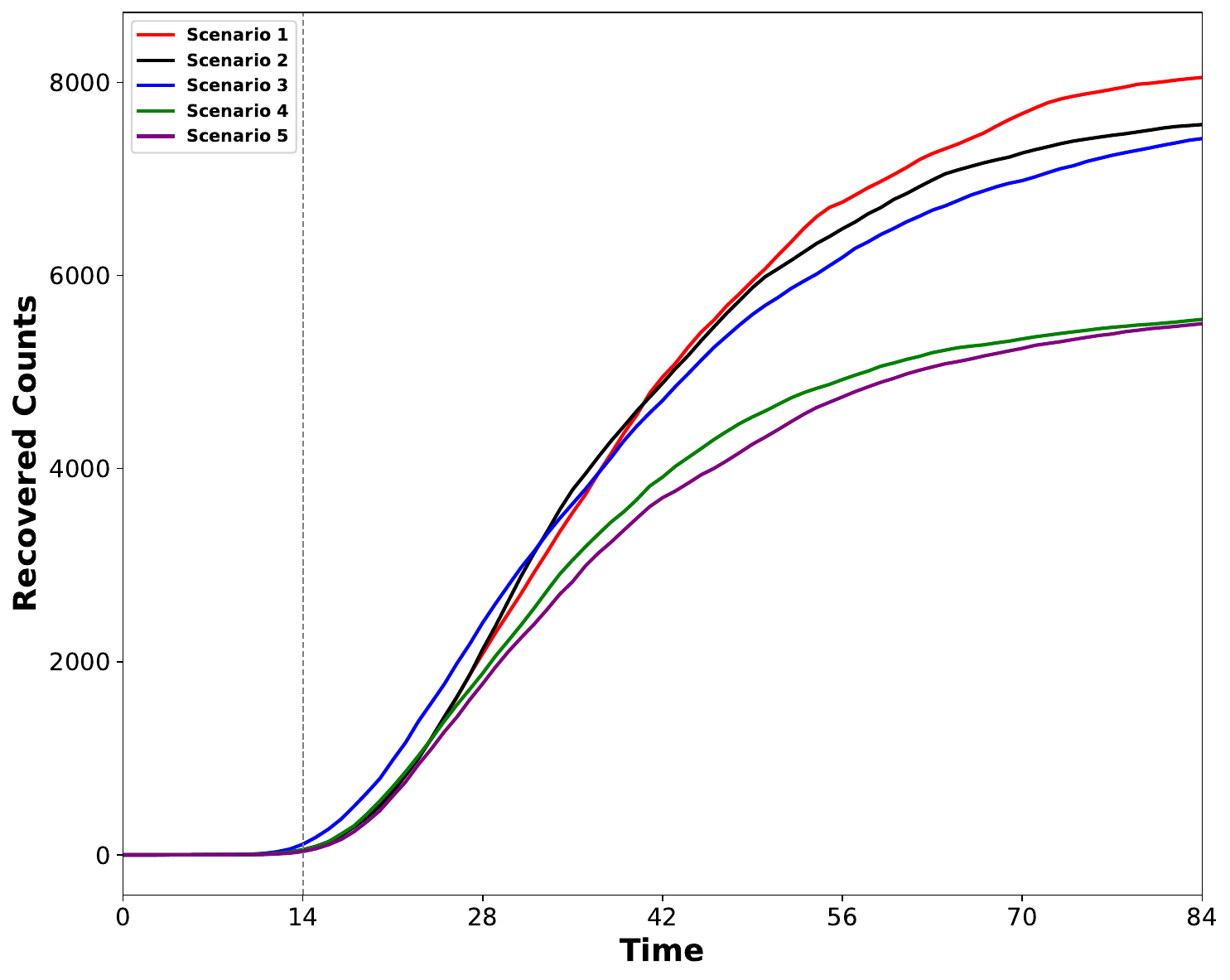}
     \caption{Recovered}
     \label{fig:reco}
     \end{center}
 \end{subfigure}
   \hspace*{\fill} %
  \caption{Median epidemic trajectories for the susceptible, exposed, infected, recovered, and deceased populations under five intervention scenarios: Scenario 1 (no restrictions), Scenario 2 (school-from-home), Scenario 3 (essential workers only), Scenario 4 (rotational shifts among essential workers with quarantine), and Scenario 5 (rotational shifts among essential workers without quarantine). Results reflect the median across 100 simulation runs, each conducted with a distinct synthetic population of 10,000 agents. The vertical reference line at day 14 indicates the start of the intervention period; all scenarios follow the same baseline contact structure before day 14, after which scenario-specific intervention rules are applied.}\label{fig:lineScenarios}
\end{figure}

Figure~\ref{fig:lineScenarios} illustrates the epidemic trajectories for susceptible, exposed, infected (symptomatic), recovered, and deaths populations across five distinct intervention scenarios, with a vertical reference line at day 14 indicating the start of scenario-specific interventions. Scenario 1 (no restrictions, red) serves as the baseline, while Scenario 2 (school-from-home, black), Scenario 3 (essential-workers-only, blue), Scenario 4 (rotational shifts among essential workers with quarantine, green), and  Scenario 5 (rotational shifts among essential workers without quarantine, purple), represent distinct structural intervention strategies.

Panels (a) and (b) present the susceptible and exposed (asymptomatic) populations. In Scenario 2 (school-from-home), the susceptible population drops significantly starting on day 16, soon after the intervention begins, indicating substantial movement of agents from the susceptible state into the exposed state during the early intervention period. This early decline should not be interpreted as intervention success; rather, it reflects unintended family transmission dynamics. Children confined at home, while parents continue external interactions through work and other necessary activities, may experience increased within-family contacts and infection risk. Although the susceptible trajectory in Scenario 2 stabilizes after approximately day 31, the initial rapid decline suggests that school-from-home alone may temporarily shift transmission into household settings when adult external contacts are not also reduced. This demonstrates a crucial insight: closing schools without reducing parental interactions outside the home may inadvertently increase family-level transmission for some period of time. Scenario 3 (essential-workers-only, blue) shows a slower reduction in susceptible individuals compared with Scenarios 1 and 2, reflecting the benefit of restricting external mobility to essential workers. However, panel (b) shows that the exposed population under Scenario 3 remains elevated for a longer period. By around day 40, the exposed count in Scenario 3 becomes comparable to Scenario 1, and after approximately day 48, it exceeds Scenario 1. This indicates prolonged vulnerability among essential workers and their families when external mobility is concentrated within this group. In contrast, the rotational shift strategies provide stronger protection. Scenario 5, which applies rotational shifts without quarantine, shows substantial improvement compared with Scenarios 1--3. By the end of the simulation, approximately 2,976 agents remain susceptible, indicating that splitting essential workers into mutually exclusive groups and alternating their workdays every 7 days substantially reduces transmission even without quarantine. However, the best strategy remains Scenario 4, where rotational shifts are combined with quarantine. Scenario 4 consistently preserves the largest susceptible population and produces the lowest exposed counts over time. By the end of the simulation, approximately 3,136 agents remain susceptible, showing additional protection beyond rotation alone. These results indicate that workforce rotation reduces sustained exposure among essential workers, while quarantining symptomatic active workers further strengthens transmission control and provides more sustained protection for this critical population.

Panel (c) aligns logically with expected epidemiological patterns: higher exposure leads directly to more symptomatic infections. Scenario 2 experiences the greatest infection counts from approximately day 28 to day 52; after day 52, Scenarios 1 and 2 show similarly high infection counts, directly reflecting their elevated exposure levels. Scenario 4 achieves the lowest infection counts over time, distinctly separating from Scenario 5 around day 28 and maintaining consistently lower counts thereafter. It is also worth noting that Scenario 5 has a lower infection burden compared with Scenarios 1--3, further showing the effectiveness of splitting essential workers into mutually exclusive groups. This sustained improvement emphasizes the benefit of structured workforce rotation, with the strongest reduction observed when rotation is combined with targeted quarantine practices. Panels (d) and (e) illustrate cumulative deaths and recoveries, consistent with the infection trajectories. Scenarios 1 and 2 show higher recoveries due to elevated infections over time, confirming that recovery counts alone should not be interpreted as intervention success. Scenario 5 shows the lower median cumulative recovery trajectory, whereas Scenario 4 produces the lowest cumulative deaths and preserves the largest susceptible population, aligning with the effectiveness of rotational workforce strategies in reducing initial exposures and symptomatic infections. Clearly, the strategy of splitting essential workers into mutually exclusive groups and promptly quarantining symptomatic individuals appears to be the most protective approach for this high-risk population, particularly during pandemics when critical societal functions must continue.

\subsection{Uncertainty Quantification Across all Scenarios }

\subsubsection{ Predictive uncertainty in cumulative recoveries}
\begin{figure}[H]
 \begin{subfigure}[t]{0.5\textwidth}
  \centering
     \includegraphics[width=\textwidth]{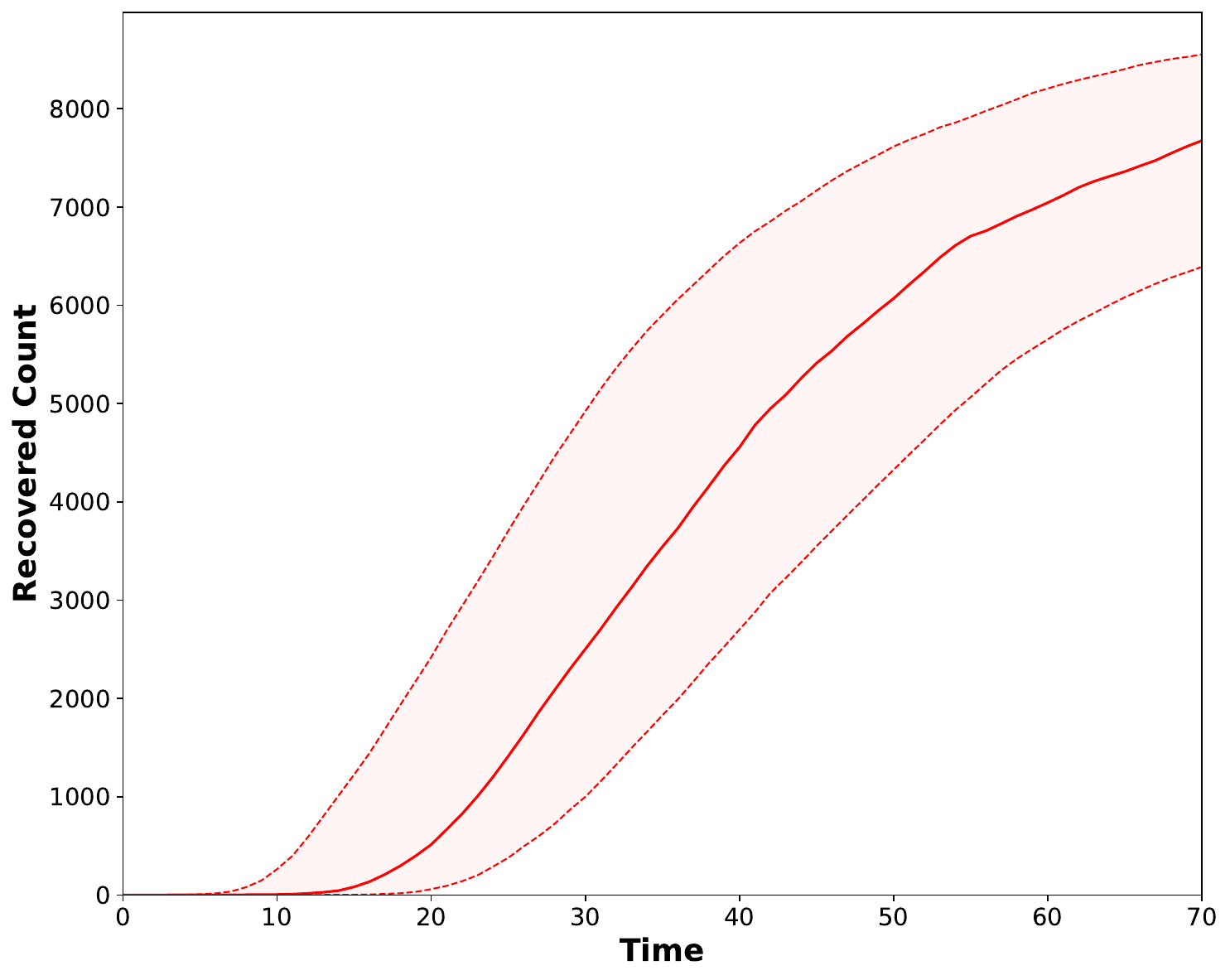}
     \caption{Scenario 1}
     \label{fig:recovNR}
 \end{subfigure}
 \hfill
 \begin{subfigure}[t]{0.5\textwidth}
  \centering
     \includegraphics[width=\textwidth]{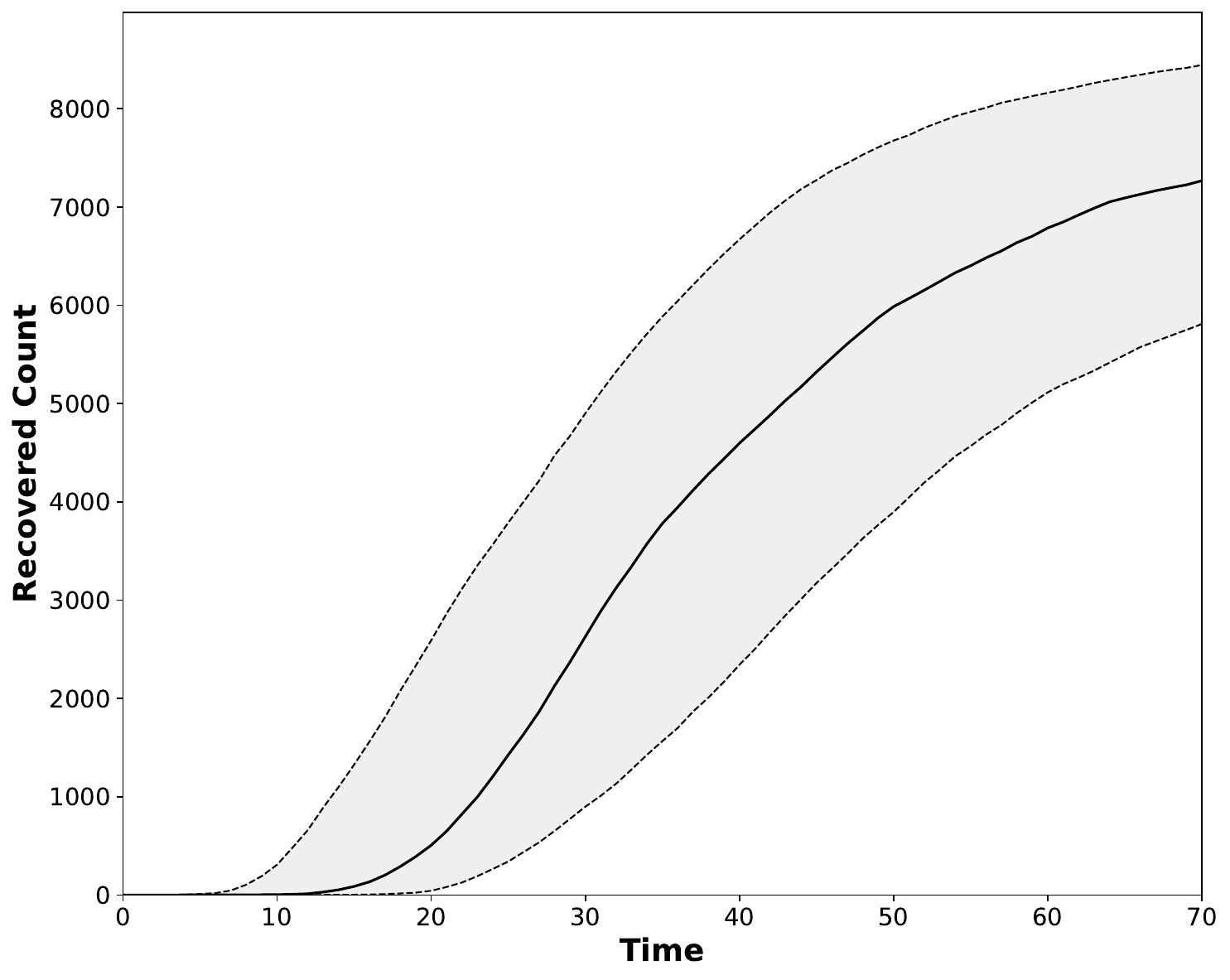}
     \caption{Scenario 2}
     \label{fig:recovSH}
 \end{subfigure}
 \begin{subfigure}[t]{0.5\textwidth}
  \centering
    \includegraphics[width=\textwidth]{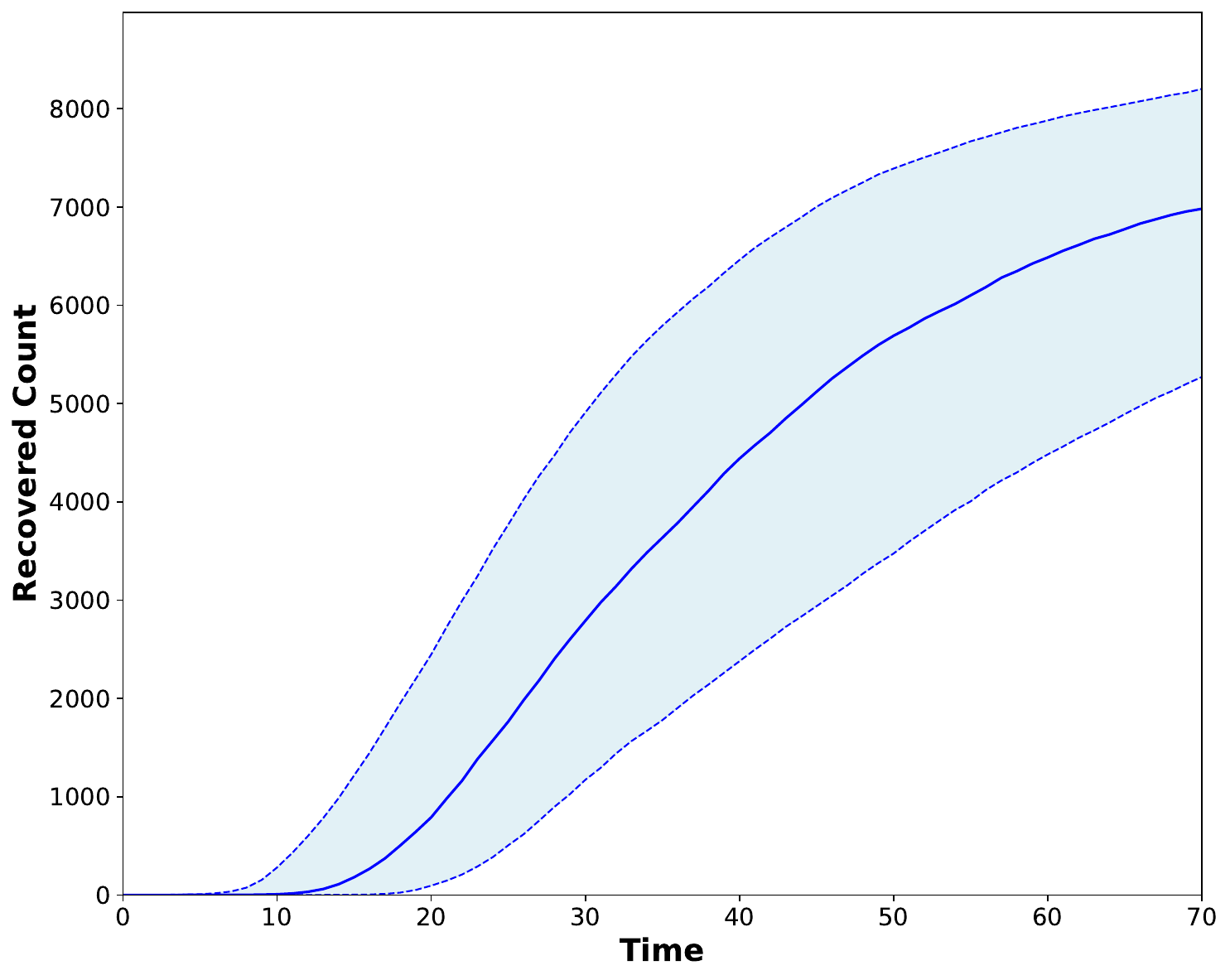}
     \caption{Scenario 3}
     \label{fig:recovEW}
 \end{subfigure}
 \hfill
 \begin{subfigure}[t]{0.5\textwidth}
  \centering
     \includegraphics[width=\textwidth]{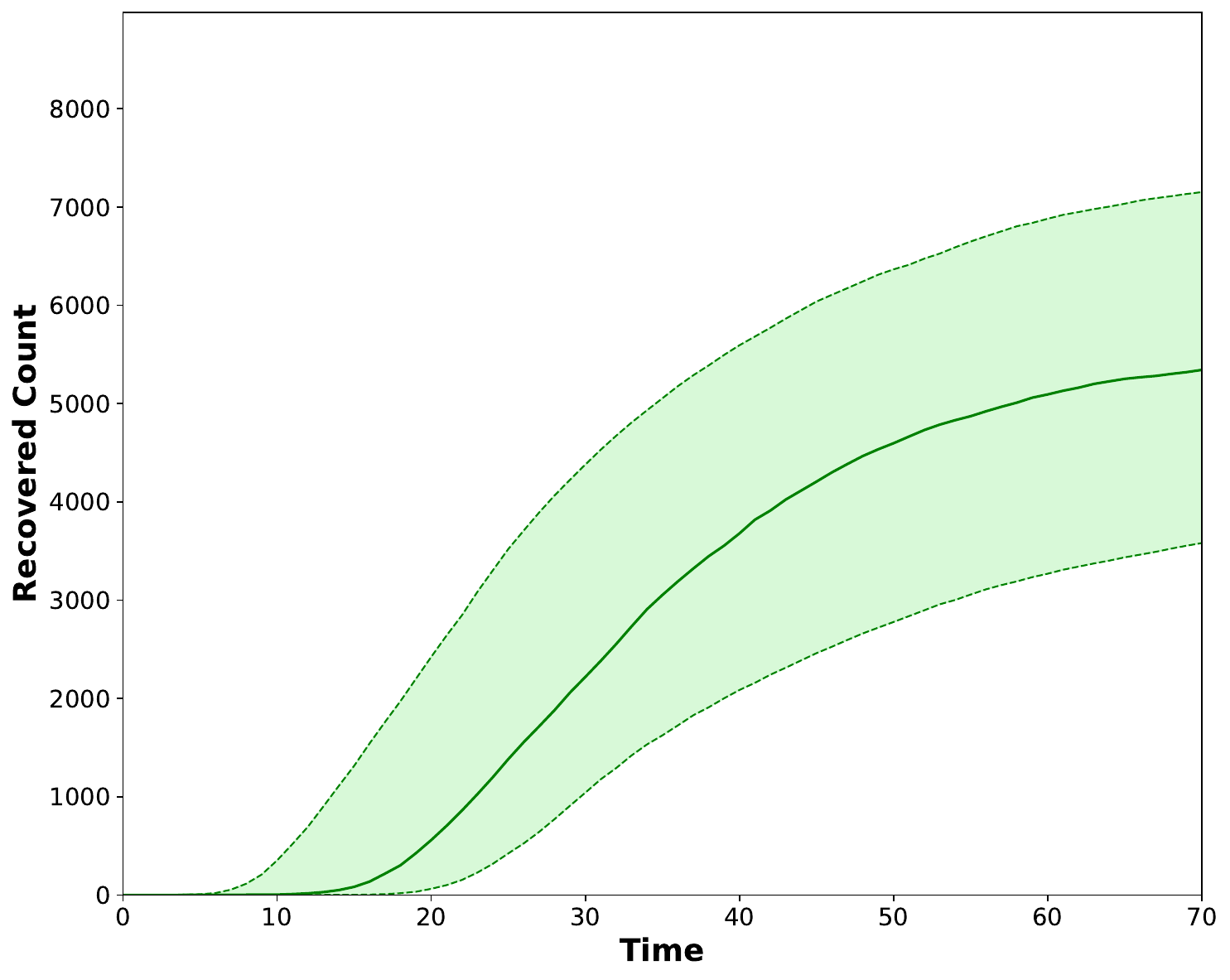}
     \caption{Scenario 4}
     \label{fig:recovSEW}
 \end{subfigure}
 
\begin{center}
\begin{subfigure}[t]{0.5\textwidth}
  \centering
  \includegraphics[width=\textwidth]{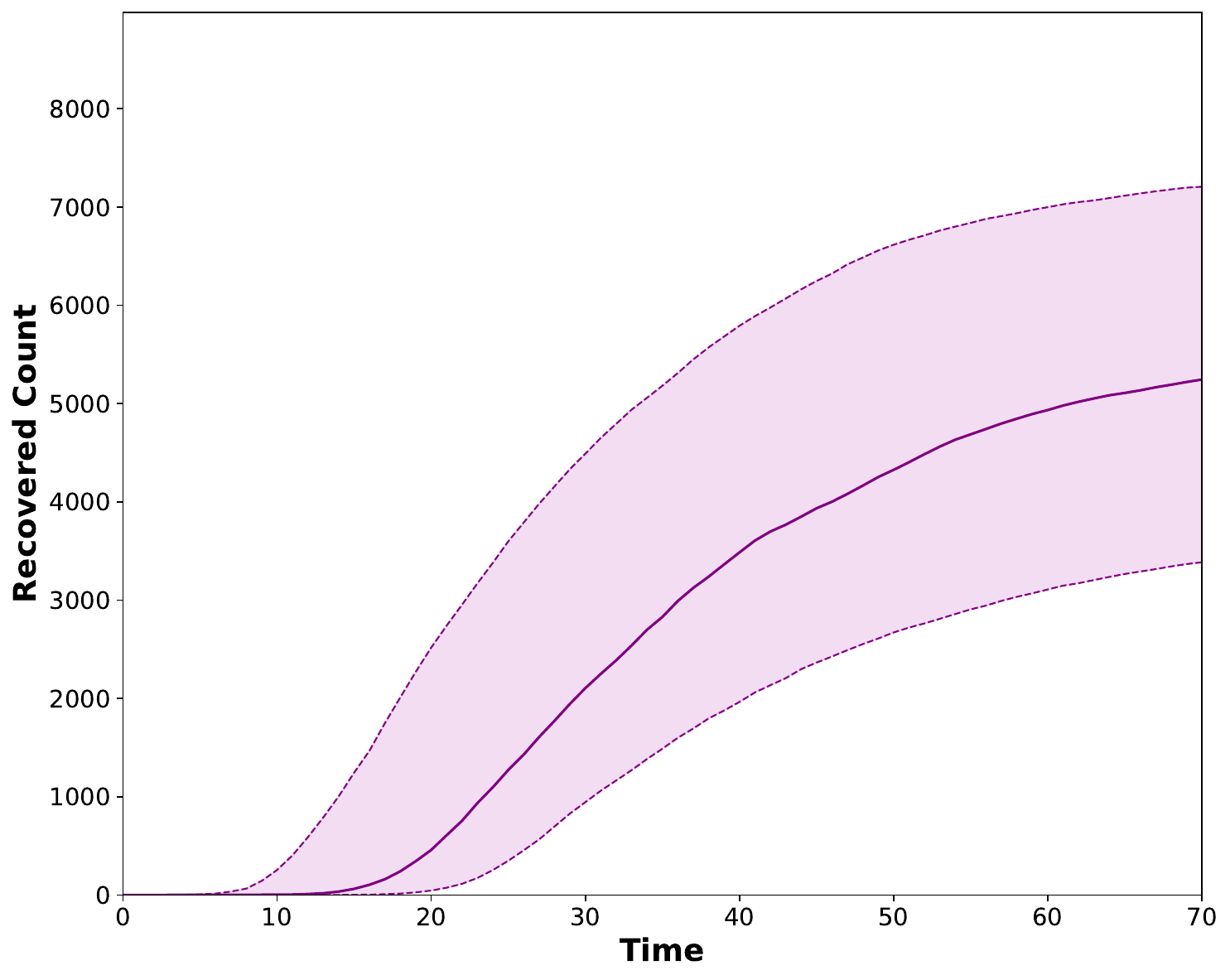}
  \caption{Scenario 5}
  \label{fig:recovSEWnoQ}
\end{subfigure}
\end{center}
  \caption{ Predictive uncertainty in recovered population across five intervention scenarios. Each panel shows the median trajectory (solid line) with the 2.5th and 97.5th percentiles (shaded region) based on 100 simulation runs. Each run used a distinct synthetic population of 10,000 agents. The quantile bands reflect uncertainty from both stochastic dynamics and parameter variation. Colors are consistent across scenarios, with each scenario represented by the same color used in earlier figures for ease of comparison.}\label{fig:recovScenarios}
\end{figure}

Figure~\ref{fig:recovScenarios} shows predictive uncertainty in cumulative recoveries for each of the five intervention scenarios. For consistency, Scenario 1 (no restrictions) is shown in red, Scenario 2 (school-from-home) in black, Scenario 3 (essential-workers-only) in blue, Scenario 4 (rotational shifts among essential workers with quarantine) in green, and Scenario 5 (rotational shifts among essential workers without quarantine) in purple.

In Scenarios 1 and 2, cumulative recoveries rise quickly, with Scenario 1 reaching the highest overall recovery count by the end of the simulation and Scenario 2 remaining slightly lower cumulatively. However, these higher recovery counts should not be interpreted as evidence of better epidemic management. Rather, they reflect widespread infections under no restriction and incomplete control under the school-from-home intervention, where transmission continues through household and adult external contacts. The broad predictive interval (2.5th--97.5th percentiles) in Scenario 1 further highlights substantial variability and unpredictability in epidemic outcomes across simulation runs under minimal restrictions. Scenario 3 displays slower growth in recoveries, reflecting fewer infections due to structural interventions that restrict contact and mobility.

Scenario 3 shows slower cumulative recovery growth than Scenarios 1 and 2, reflecting fewer infection-driven recoveries under the essential-workers-only intervention. However, its predictive interval is not consistently narrower than those of Scenarios 1 and 2 across the full simulation period, indicating that reduced cumulative recovery growth should be interpreted separately from recovery-trajectory uncertainty. Scenarios 4 and 5 produce lower cumulative recovery trajectories than Scenarios 1--3, reflecting the effectiveness of limiting overall infections through strategic workforce splitting. Among the two rotational workforce strategies, Scenario 5 shows the lower median cumulative recovery trajectory, whereas Scenario 4 shows a narrower predictive interval, indicating greater consistency in recovery outcomes under the quarantine-based rotation strategy. Thus, by dividing essential workers into mutually exclusive groups alternating every 7 days and promptly quarantining symptomatic individuals, the quarantine-based rotational intervention reduces transmission opportunities and improves the stability of recovery outcomes within this high-risk population.

Moreover, Scenarios 4 and 5 show more stable recovery trajectories compared with Scenarios 1--3, indicating more controlled epidemic dynamics across simulation runs. This is important because Scenario 5 demonstrates that workforce rotation alone provides meaningful benefit even without quarantine, while Scenario 4 shows that adding quarantine further strengthens this protection. These findings demonstrate the value of explicitly modeling interventions tailored specifically to essential workers, a critical population during pandemics when societal functions must continue. By structurally modifying interactions rather than relying solely on generalized parameter adjustments, this targeted approach substantially reduces epidemic uncertainty and stabilizes long-term recovery outcomes, offering critical insights for high-risk population management.

To further illustrate how the interventions alter epidemic timing, we examine the inflection points in the recovery trajectories. The inflection point is defined as the time at which the rate of cumulative recoveries begins to decline, indicating a slowdown in the number of new recoveries per time step. Because recoveries occur after infection, this transition reflects changes in the preceding infection process as well as the effect of intervention strategies on epidemic progression. In Scenario 1, the inflection point in the median recovery trajectory occurs around day 27, with a mean of 30.71 $\pm$ 6.62 days across simulation runs, reflecting an uncontrolled surge in infections followed by widespread recovery. Scenario 2 shows a similar but slightly later inflection point in the median recovery trajectory around day 28, with a mean of 29.75 $\pm$ 6.79 days across simulation runs, suggesting that school-from-home alone does not substantially shift the recovery trajectory into an earlier slowing phase.

By contrast, Scenarios 3, 4, and 5 show earlier recovery inflection patterns. For Scenario 3, the inflection point in the median recovery trajectory occurs around day 23, with a mean of 25.37 $\pm$ 5.16 days, indicating that restricting mobility to essential workers reduces the pace of infection-driven recoveries compared with Scenarios 1 and 2. For Scenario 5, the inflection point in the median recovery trajectory also occurs around day 23, with a mean of 25.19 $\pm$ 4.21 days, showing that rotational splitting of essential workers remains effective even without quarantine. For Scenario 4, the inflection point in the median recovery trajectory occurs around day 25, with the lowest mean recovery inflection day, 24.36 $\pm$ 4.57 days, indicating a similarly early and stable transition under rotational shifts with quarantine. These results show that stronger structural interventions move the recovery trajectory into a slower accumulation phase earlier than Scenarios 1 and 2, with Scenarios 4 and 5 further supporting the benefit of rotational splitting among essential workers.

\begin{table}[htbp]

\centering
\caption{Inflection points for cumulative recovery and death curves across simulation scenarios.}
\resizebox{\textwidth}{!}{
\begin{tabular}{lcccccc}
\toprule
& \multicolumn{3}{c}{Recovery inflection day} & \multicolumn{3}{c}{Death inflection day} \\
\cmidrule(lr){2-4} \cmidrule(lr){5-7}
Scenario 
& Inflection point & Mean $\pm$ SD & 95\% interval
& Inflection point & Mean $\pm$ SD & 95\% interval \\
\midrule
Scenario 1: No Restriction 
& 27 & 30.71 $\pm$ 6.62 & 20.00--42.00
& 35 & 35.31 $\pm$ 7.54 & 22.00--47.52 \\

Scenario 2: School From Home 
& 28 & 29.75 $\pm$ 6.79 & 18.00--43.52
& 32 & 34.41 $\pm$ 6.79 & 23.00--46.00 \\

Scenario 3: Essential Workers Only 
& 23 & 25.37 $\pm$ 5.16 & 17.00--34.00
& 38 & 34.56 $\pm$ 8.42 & 21.00--53.57 \\

Scenario 4: Split Essential Workers + Quarantine
& 25 & 24.36 $\pm$ 4.57 & 17.00--32.00
& 32 & 30.48 $\pm$ 5.48 & 21.00--40.00 \\

Scenario 5: Split Essential Workers without Quarantine
& 23 & 25.19 $\pm$ 4.21 & 17.48--32.52
& 32 & 30.94 $\pm$ 5.67 & 21.48--41.52 \\
\bottomrule
\end{tabular}}\label{Table:comaprison}
\vspace{0.3em}
\parbox{\textwidth}{\scriptsize\raggedright
Note: Inflection days were computed numerically from the cumulative recovery and death trajectories using the finite-difference rate-decline criterion described in Section 4.17. The inflection point column reports the value obtained by applying this criterion to the median trajectory shown in Figures~\ref{fig:recovScenarios} and~\ref{fig:deathsScenarios}, while the mean $\pm$ SD and 95\% interval summarize the inflection days obtained by applying the same criterion separately across 100 simulation runs.
}
\end{table}

\subsubsection{ Predictive uncertainty in cumulative deaths}
\begin{figure}[H]
 \begin{subfigure}[t]{0.5\textwidth}
  \centering
     \includegraphics[width=\textwidth]{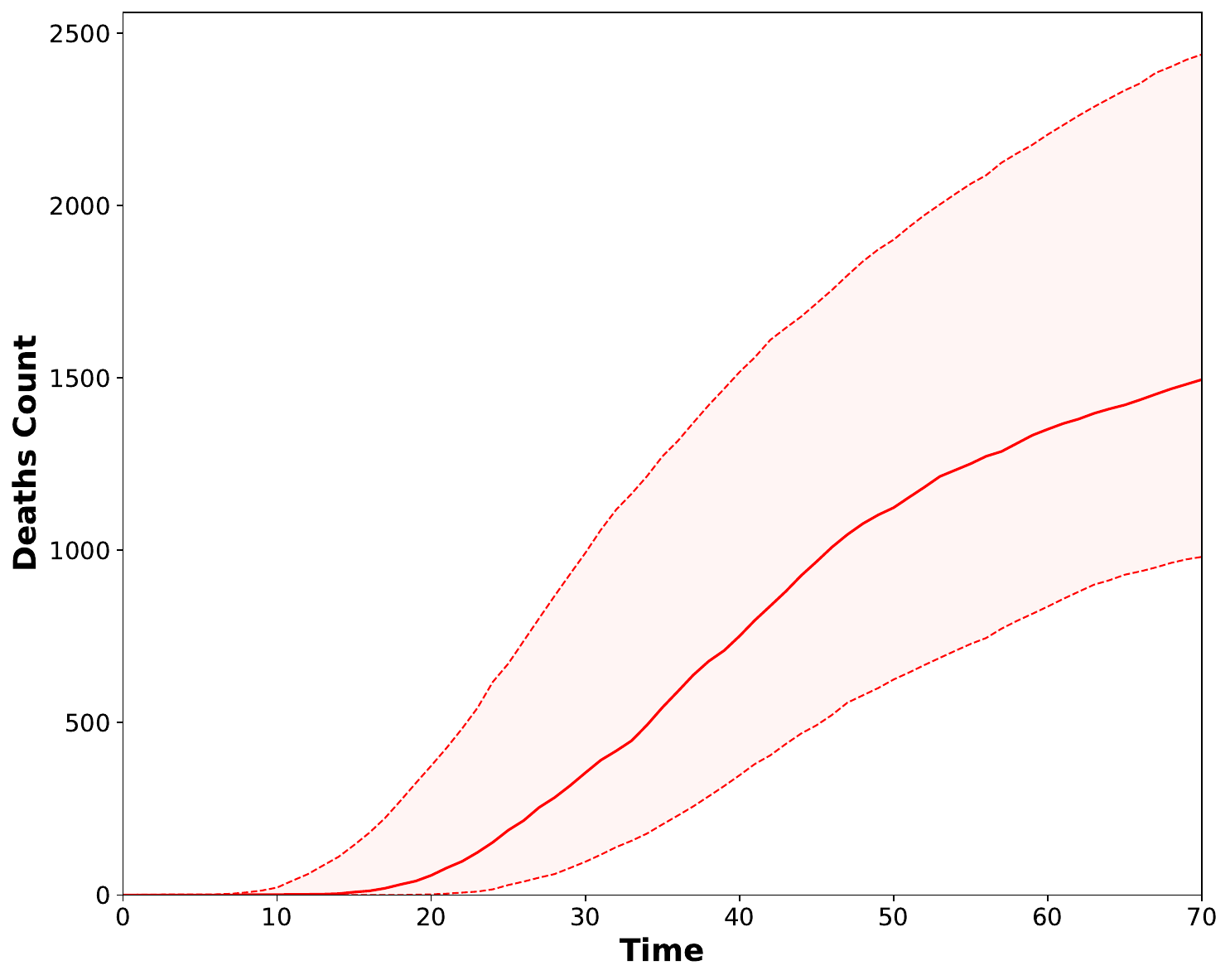}
     \caption{Scenario 1}
     \label{fig:deathsNR}
 \end{subfigure}
 \hfill
 \begin{subfigure}[t]{0.5\textwidth}
  \centering
     \includegraphics[width=\textwidth]{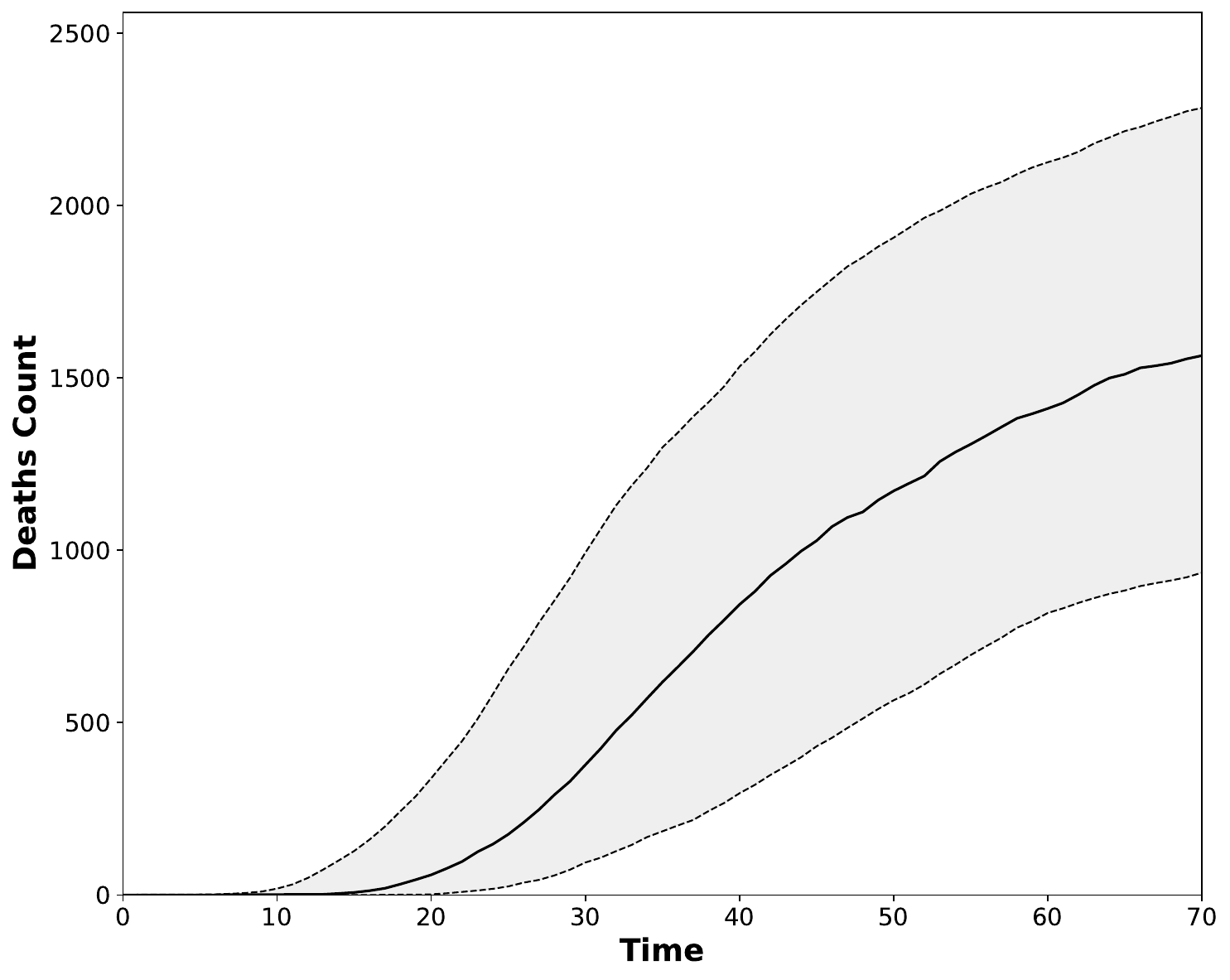}
     \caption{Scenario 2}
     \label{fig:deathsSH}
 \end{subfigure}
 \begin{subfigure}[t]{0.5\textwidth}
  \centering
    \includegraphics[width=\textwidth]{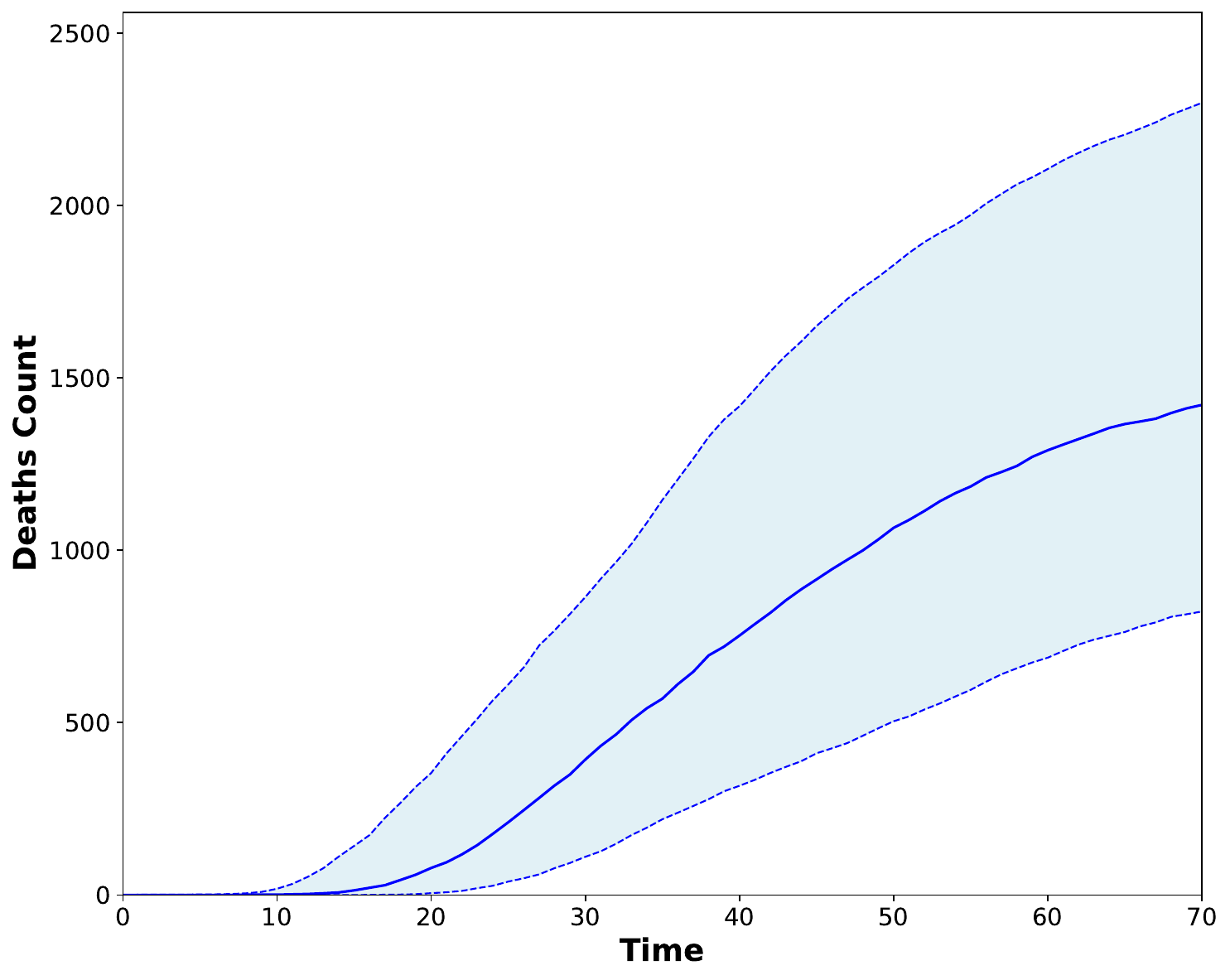}
     \caption{Scenario 3}
     \label{fig:deathsEW}
 \end{subfigure}
 \hfill
 \begin{subfigure}[t]{0.5\textwidth}
  \centering
     \includegraphics[width=\textwidth]{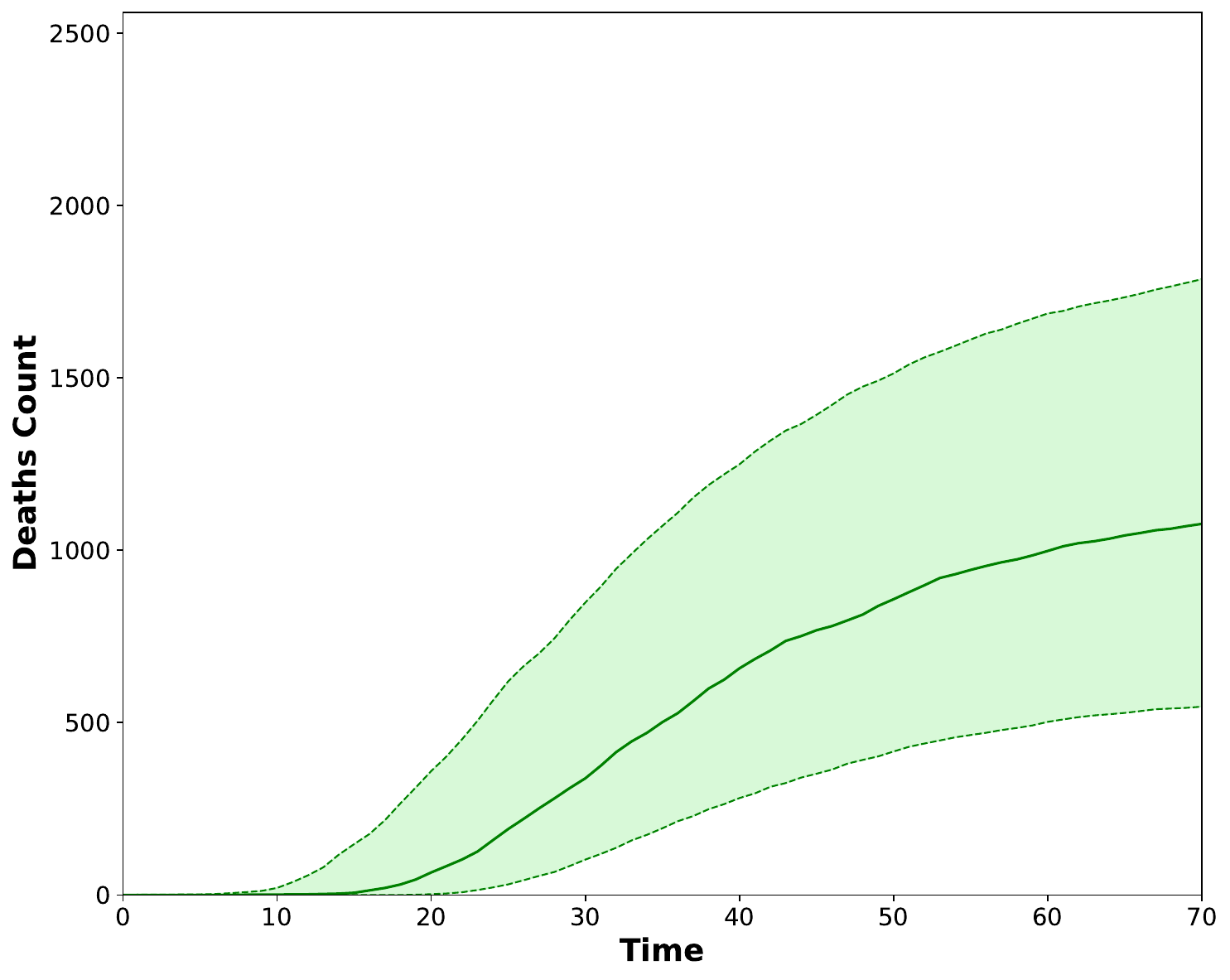}
     \caption{Scenario 4}
     \label{fig:deathsSEW}
 \end{subfigure}

 \begin{center}
      \begin{subfigure}[t]{0.5\textwidth}
  \centering
     \includegraphics[width=\textwidth]{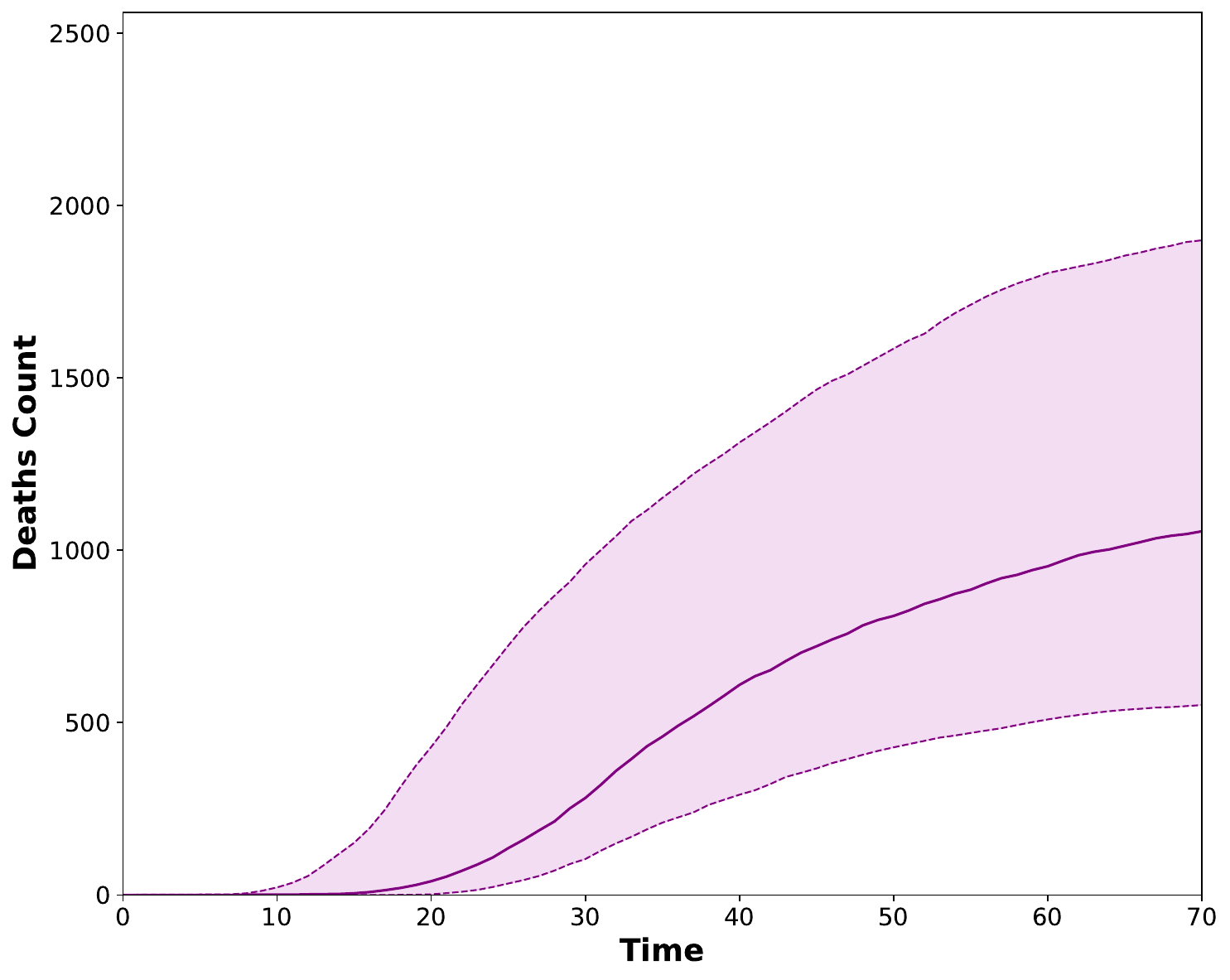}
     \caption{Scenario 5}
     \label{fig:deathsSEWnoQ}
 \end{subfigure}
 \end{center}
   \caption{ Predictive uncertainty in cumulative deaths under five intervention scenarios. Each panel presents the median trajectory (solid line) and the 2.5th and 97.5th percentiles (shaded region), computed from 100 simulation runs using distinct synthetic populations of 10,000 agents. The uncertainty bands incorporate variability due to both stochastic interactions and parameter uncertainty. Scenario colors are consistent with earlier figures for ease of comparison.}\label{fig:deathsScenarios}
\end{figure}

Similarly, Figure~\ref{fig:deathsScenarios} examines predictive uncertainty in cumulative deaths under each intervention scenario, using the same consistent scenario color scheme as before: Scenario 1 (no restrictions, red), Scenario 2 (school-from-home, black), Scenario 3 (essential-workers-only, blue), Scenario 4 (rotational shifts among essential workers with quarantine, green), and Scenario 5 (rotational shifts among essential workers without quarantine, purple).

In Scenario 1, cumulative deaths increase rapidly, resulting in the highest death burden and one of the broadest predictive intervals among all scenarios. The extensive variability in this scenario reflects the inherent unpredictability of epidemics under no restriction, where unchecked transmission can lead to substantially different mortality trajectories across simulation runs. Compared with the recovery trajectories, the death trajectories generally show greater uncertainty, which is expected because deaths occur downstream of infection and depend on stochastic progression from exposure to symptomatic infection and from infection to mortality. Introducing structural interventions reduces cumulative deaths and improves the overall stability of mortality outcomes. Scenario 3 produces lower cumulative deaths compared with Scenario 1, while Scenario 2 shows more limited improvement, indicating that school-from-home alone does not fully control death burden when household transmission pathways remain active.

Scenarios 4 and 5 show the greatest reduction in cumulative deaths, with Scenario 4 remaining the most effective. Scenario 5 demonstrates that rotational workforce splitting alone reduces mortality compared with Scenarios 1--3, while Scenario 4 shows the additional benefit of combining rotation with timely quarantine of symptomatic active workers. Through strategic rotation of essential worker groups and prompt quarantine, Scenario 4 reduces sustained workplace exposure and limits secondary transmission among this critical high-risk group. Although uncertainty remains visible in the death trajectories, the lower cumulative deaths in Scenarios 4 and 5 indicate more controlled mortality outcomes under targeted structural interventions. These findings demonstrate that explicit modifications of interaction patterns can significantly improve outcomes for essential workers and their families, even under varying population compositions. They also highlight that interventions designed specifically for essential workers, who are critical to maintaining societal functions during pandemics, can substantially reduce epidemic severity, minimize mortality, and improve the stability of epidemic outcomes.

To better evaluate intervention impacts on mortality dynamics, we also examine the inflection points of cumulative death trajectories. The inflection point represents the time when the rate of cumulative deaths begins to slow down, marking a crucial turning point where new deaths per time step start to decrease either due to intervention effects or the natural progression of the epidemic. In Scenario 1, the death inflection point occurs around day 35, with a mean of 35.31 $\pm$ 7.54 days across simulation runs, reflecting a prolonged period of mortality accumulation under no restriction. Scenario 3 shows the latest inflection point from the median trajectory, around day 38, and also has the widest 95\% interval, suggesting greater variability in the timing of mortality slowing under the essential-workers-only intervention. Scenario 2 has an inflection point around day 32, but its mean inflection timing remains closer to Scenarios 1 and 3 than to the rotational shift strategies, indicating that school-from-home modifies transmission patterns but does not stabilize mortality dynamics as strongly as the rotational approaches.

Scenarios 4 and 5 both show death inflection points around day 32 from the median trajectories, but the summaries across the 100 simulation runs indicate earlier and more stable mortality transitions than Scenarios 1--3. Scenario 4 has the lowest mean death inflection time, 30.48 $\pm$ 5.48 days, while Scenario 5 is very close at 30.94 $\pm$ 5.67 days. This suggests that splitting essential workers into mutually exclusive rotating groups helps move the death trajectory into a slower accumulation phase earlier in the epidemic course. Importantly, Scenario 5 shows that this improvement remains evident even without quarantine, supporting the effectiveness of the rotational shift structure itself. However, the slightly stronger performance of Scenario 4 indicates that adding prompt quarantine of symptomatic active workers provides additional protection and further stabilizes mortality outcomes.

\subsection{Statistical validation and sensitivity analysis}
To provide a formal numerical evaluation of the cumulative recovery and death outcomes shown in Figures~\ref{fig:recovScenarios} and \ref{fig:deathsScenarios}, we summarized the final cumulative values across the 100 simulation runs. As shown in Table~\ref{tab:outcome}, Scenarios 4 and 5 produced substantially lower final cumulative recoveries and deaths than Scenarios 1--3. Since cumulative recoveries reflect individuals who were previously infected, the lower final recovery counts under Scenarios 4 and 5 should be interpreted as evidence of reduced infection spread, rather than poorer recovery performance. Scenario 4 produced the lowest final cumulative deaths, followed closely by Scenario 5, indicating that rotational splitting of essential workers strongly reduces mortality outcomes, with additional descriptive benefit when quarantine is included.

\begin{table}[htbp]

\centering
\caption{Mean and standard deviation of final cumulative recovery and death outcomes across simulation scenarios.}
\resizebox{\textwidth}{!}{
\begin{tabular}{lcc}
\toprule
Scenario & Final cumulative recovered & Final cumulative deaths \\
\midrule
Scenario 1: No Restriction 
& 8202.88 $\pm$ 465.64 
& 1745.26 $\pm$ 445.24 \\

Scenario 2: School From Home 
& 7820.77 $\pm$ 506.65 
& 1763.69 $\pm$ 413.81 \\

Scenario 3: Essential Workers Only 
& 7681.28 $\pm$ 570.50 
& 1661.42 $\pm$ 427.83 \\

Scenario 4: Split Essential Workers + Quarantine
& 5624.76 $\pm$ 991.37 
& 1214.79 $\pm$ 371.25 \\

Scenario 5: Split Essential Workers without Quarantine
& 5650.54 $\pm$ 957.98 
& 1259.52 $\pm$ 428.35 \\
\bottomrule
\end{tabular}}\label{tab:outcome}
\vspace{0.3em}
\parbox{\textwidth}{\scriptsize\raggedright
\textit{Note:} Values are reported as mean $\pm$ SD across 100 simulation runs. Final cumulative recovered and death counts were computed at the final simulation time point (day 120).
}
\end{table}

To determine whether scenario-specific outcome distributions differed statistically, we conducted pairwise two-sided Welch t-tests, allowing for unequal variances, with statistical significance assessed at $\alpha=0.05$. The results in Table~\ref{tab:welch} show that Scenarios 4 and 5 differed significantly from Scenarios 1--3 for both final cumulative recoveries and final cumulative deaths. In contrast, the comparison between Scenarios 4 and 5 was not statistically significant for either final cumulative recovered or final cumulative deaths, suggesting that the rotational shift strategy without quarantine performs similarly to the rotational shift strategy with quarantine in terms of final cumulative outcomes. The Welch tests in Table~\ref{tab:welch} for inflection days further show that several scenario differences in curve timing are statistically supported, particularly for comparisons involving Scenarios 4 and 5. However, these inflection results are best interpreted together with the cumulative trajectories and final outcome summaries, since inflection timing alone does not fully characterize intervention effectiveness.

\begin{table}[htbp]
\centering
\caption{Pairwise Welch t-test results for final outcomes and inflection days across simulation scenarios.}
\resizebox{\textwidth}{!}{
\begin{tabular}{lcccc}
\toprule
& \multicolumn{1}{c}{Final recovered} & \multicolumn{1}{c}{Final deaths} & \multicolumn{1}{c}{Recovery inflection day} & \multicolumn{1}{c}{Death inflection day} \\
\cmidrule(lr){2-2} \cmidrule(lr){3-3} \cmidrule(lr){4-4} \cmidrule(lr){5-5}
Comparison & Welch $p$ & Welch $p$ & Welch $p$ & Welch $p$ \\
\midrule
S1 vs S2 & $< 0.001$ & 0.762 & 0.313 & 0.376 \\
S1 vs S3 & $< 0.001$ & 0.176 & $< 0.001$ & 0.508 \\
S1 vs S4 & $< 0.001$ & $< 0.001$ & $< 0.001$ & $< 0.001$ \\
S1 vs S5 & $< 0.001$ & $< 0.001$ & $< 0.001$ & $< 0.001$ \\
S2 vs S3 & 0.069 & 0.087 & $< 0.001$ & 0.890 \\
S2 vs S4 & $< 0.001$ & $< 0.001$ & $< 0.001$ & $< 0.001$ \\
S2 vs S5 & $< 0.001$ & $< 0.001$ & $< 0.001$ & $< 0.001$ \\
S3 vs S4 & $< 0.001$ & $< 0.001$ & 0.144 & $< 0.001$ \\
S3 vs S5 & $< 0.001$ & $< 0.001$ & 0.787 & $< 0.001$ \\
S4 vs S5 & 0.852 & 0.431 & 0.183 & 0.560 \\
\bottomrule
\end{tabular}}\label{tab:welch}
\vspace{0.3em}
\parbox{\textwidth}{\scriptsize\raggedright
\textit{Note:} Welch $p$ values are from two-sided Welch t-tests comparing scenario distributions while allowing for unequal variances. S1 = Scenario 1, S2 = Scenario 2, S3 = Scenario 3, S4 = Scenario 4, and S5 = Scenario 5.
}
\end{table}

Finally, Table~\ref{tab:mutual} summarizes the mutual information ranking of input effects on selected model outputs. The results show that $\tau_{\text{incubation}}$ is consistently among the strongest contributors to day-30 recovery and death outcomes, indicating that early epidemic dynamics are strongly influenced by the incubation period. For final cumulative outcomes, $P_{I\to R}$ and $P_{I\to D}$ also appear among the leading contributors across several scenarios, reflecting their direct roles in determining recovery and mortality accumulation over time. Seed index appears among the top-ranked inputs only in limited cases, suggesting that variation in the sampled epidemiological parameters generally contributes more strongly to output variability than population-seed effects alone. Overall, these statistical comparisons and sensitivity rankings provide additional quantitative support for the scenario-based findings observed in the cumulative recovery and death trajectories.

\begin{table}[htbp]
\centering
\caption{Mutual information ranking of input effects on selected model outputs.}
\resizebox{\textwidth}{!}{
\begin{tabular}{llccc}
\toprule
Scenario & Output & Rank 1 & Rank 2 & Rank 3 \\
\midrule
S1 & Final recovered & $P_{I\to R}$ (0.567) & $P_{I\to D}$ (0.264) & $\tau_{\text{incubation}}$ (0.063) \\
S1 & Final deaths & $P_{I\to R}$ (0.454) & $P_{I\to D}$ (0.331) & $\tau_{\text{incubation}}$ (0.036) \\
S1 & Recovered at day 30 & $\tau_{\text{incubation}}$ (0.854) & $\beta$ (0.044) & $P_{I\to D}$ (0.031) \\
S1 & Deaths at day 30 & $\tau_{\text{incubation}}$ (0.906) & $P_{I\to D}$ (0.141) & $\beta$ (0.055) \\

S2 & Final recovered & $P_{I\to R}$ (0.207) & $\beta$ (0.183) & $P_{E\to R}$ (0.165) \\
S2 & Final deaths & $P_{I\to D}$ (0.263) & $P_{I\to R}$ (0.258) & $\tau_{\text{incubation}}$ (0.115) \\
S2 & Recovered at day 30 & $\tau_{\text{incubation}}$ (0.883) & $\beta$ (0.175) & $P_{E\to I}$ (0.087) \\
S2 & Deaths at day 30 & $\tau_{\text{incubation}}$ (0.781) & $P_{I\to D}$ (0.249) & $P_{E\to R}$ (0.093) \\

S3 & Final recovered & $\tau_{\text{incubation}}$ (0.364) & $P_{E\to R}$ (0.091) & $P_{I\to R}$ (0.085) \\
S3 & Final deaths & $P_{I\to R}$ (0.387) & $P_{I\to D}$ (0.237) & $\tau_{\text{incubation}}$ (0.107) \\
S3 & Recovered at day 30 & $\tau_{\text{incubation}}$ (0.701) & $\beta$ (0.098) & $P_{I\to D}$ (0.090) \\
S3 & Deaths at day 30 & $\tau_{\text{incubation}}$ (0.590) & $P_{I\to D}$ (0.107) & $P_{E\to I}$ (0.106) \\

S4 & Final recovered & $\tau_{\text{incubation}}$ (0.297) & $\beta$ (0.144) & $P_{I\to D}$ (0.003) \\
S4 & Final deaths & $P_{I\to R}$ (0.130) & $P_{E\to I}$ (0.091) & $\tau_{\text{incubation}}$ (0.078) \\
S4 & Recovered at day 30 & $\tau_{\text{incubation}}$ (0.673) & $\beta$ (0.083) & $P_{E\to I}$ (0.000) \\
S4 & Deaths at day 30 & $\tau_{\text{incubation}}$ (0.706) & $\beta$ (0.102) & $P_{E\to R}$ (0.054) \\

S5 & Final recovered & $\tau_{\text{incubation}}$ (0.294) & $\beta$ (0.286) & Seed index (0.157) \\
S5 & Final deaths & $P_{I\to R}$ (0.447) & $P_{E\to R}$ (0.217) & $\tau_{\text{incubation}}$ (0.170) \\
S5 & Recovered at day 30 & $\tau_{\text{incubation}}$ (0.815) & $P_{I\to R}$ (0.091) & $\beta$ (0.058) \\
S5 & Deaths at day 30 & $\tau_{\text{incubation}}$ (0.515) & $P_{I\to D}$ (0.131) & $\beta$ (0.072) \\
\bottomrule
\end{tabular}}\label{tab:mutual}
\vspace{0.3em}
\parbox{\textwidth}{\scriptsize\raggedright
\textit{Note:} Values in parentheses are mutual information scores. Larger values indicate stronger association between the input feature and the model output. Only the top three ranked inputs are shown for each scenario-output combination. Day 30 outcomes were included to assess input effects at an intermediate stage of the epidemic, while final recovered and final deaths summarize end-of-simulation outcomes. S1 = Scenario 1, S2 = Scenario 2, S3 = Scenario 3, S4 = Scenario 4, and S5 = Scenario 5.
}
\end{table}

\section{Conclusion}\label{conclude}

Essential workers play a critical societal role but remain disproportionately vulnerable during pandemics due to sustained high-risk interactions in occupational and community environments. To address this significant challenge, this study presents SAFE-ABM (Structured Agent-Based Framework for Essential Workers), a novel stochastic agent-based modeling framework explicitly designed to evaluate targeted intervention strategies, with an emphasis on protecting essential workers. Unlike conventional ABM approaches that often assume homogeneous mixing or simplify workforce structures, our model captures structured social interactions across families, schools, and workplaces. This detailed representation allows precise evaluation of policies such as school closures, mobility restrictions, essential-worker-only restrictions, and targeted workforce rotation among essential workers, both with and without quarantine.

A significant innovation of our framework is its integration of prior-informed uncertainty quantification, enabling systematic exploration of stochastic variability and parameter uncertainty inherent in disease transmission dynamics. Through these prior-informed simulations, our analysis quantifies intervention effectiveness probabilistically while explicitly accounting for uncertainty across simulation runs. The scenario analysis demonstrates that rotational splitting of essential workers substantially reduces transmission, even in the absence of quarantine. This is shown by Scenario 5, where splitting essential workers into mutually exclusive rotating groups lowers infection burden, cumulative recoveries, and cumulative deaths compared with Scenarios 1--3, as shown in Figure~\ref{fig:lineScenarios} and Table~\ref{tab:outcome}. When rotational splitting is combined with quarantine enforcement for symptomatic active workers, as in Scenario 4, the model produces the strongest improvement, with the lowest cumulative deaths and the largest remaining susceptible population. These findings indicate that the rotational structure itself provides meaningful protection, while quarantine adds further benefit by reducing continued transmission from symptomatic active workers. In addition, the inflection-point analysis of cumulative recoveries and deaths in Table~\ref{Table:comaprison} shows that structured interventions alter the timing and stability of epidemic slowing, supporting a more controlled epidemic trajectory under the rotational shift strategies.

The formal statistical comparisons further support the scenario-based findings. The two-sided Welch tests in Table~\ref{tab:welch} show that Scenarios 4 and 5 differ significantly from Scenarios 1--3 for final cumulative recoveries and final cumulative deaths, confirming that the rotational shift strategies produce measurable reductions in epidemic outcomes. In contrast, Scenario 4 and Scenario 5 are not statistically different from each other for final cumulative recoveries or deaths, indicating that rotational splitting without quarantine performs comparably to rotational splitting with quarantine in the final outcome distributions, although Scenario 4 remains descriptively lower for deaths. The inflection-day comparisons also show statistically supported differences in curve timing for several scenario pairs, but these timing results are best interpreted alongside the cumulative trajectories and final outcome summaries rather than as a standalone ranking of intervention effectiveness. The mutual information results in Table~\ref{tab:mutual} provide additional sensitivity evidence by identifying the inputs most strongly associated with variation in model outputs. $\tau_{\text{incubation}}$ is consistently influential for day-30 recovery and death outcomes, while $P_{I\to R}$ and $P_{I\to D}$ contribute strongly to several final cumulative outcomes. Seed index appears among the leading inputs only in limited cases, suggesting that parameter uncertainty contributes more consistently to output variability than population-seed effects alone.

While exploratory simulations confirm epidemiological plausibility, our current framework has not yet undergone calibration to empirical data. Our immediate focus was methodological rigor, structural validation, scenario analysis, statistical comparison, and sensitivity ranking rather than predictive accuracy. Formal model calibration to real-world epidemiological data thus constitutes an essential next step. Future research will integrate empirical datasets, refining parameter estimates to enhance predictive reliability and facilitate real-time evaluation of intervention strategies. In applying SAFE-ABM to a specific pandemic, posterior parameter distributions from calibrated epidemic models may also be incorporated so that parameter draws reflect the target disease and population more directly. However, this transfer requires careful consideration because infection-rate parameters estimated from population-level models may depend on population size, contact structure, and scaling assumptions. Appropriate transformations may therefore be needed before such posterior distributions can be used within an agent-based framework. Future work may also consider Bayesian parameter estimation directly within SAFE-ABM when disease parameters can reasonably be treated as stable over the study period. In practice, however, effective transmission patterns often change over time as governments and public health agencies introduce interventions that alter contact behavior and infection risk. Capturing these time-varying effects would require explicit information on the timing and nature of policy changes. Ultimately, our modeling framework, combining detailed agent interactions with prior-informed uncertainty quantification, formal scenario comparison, and input-output sensitivity ranking, provides a foundation for data-informed intervention planning. Our approach offers valuable guidance for equitable and effective public health responses in current and future pandemics by explicitly emphasizing the critical yet often overlooked role of essential workers.

\section*{Model documentation}

The SAFE-ABM model code and reproducibility materials are available through a private CoMSES Net review archive: \url{https://www.comses.net/codebase-release/3200bb24-c060-47c5-a5b5-0d45415c4279/}. The archive contains the source code, scenario-specific simulation scripts, comparison-analysis scripts, saved derived outputs, README files, software requirements, and reproducibility notes needed to review and reproduce the main results.

The archive is organized by intervention scenario, including Scenario 1 (no restrictions), Scenario 2 (school-from-home), Scenario 3 (essential-workers-only), Scenario 4 (rotational shifts among essential workers with quarantine), and Scenario 5 (rotational shifts among essential workers without quarantine). It also includes scripts used to reproduce the median epidemic trajectories, predictive-uncertainty plots, cumulative recovery and death summaries, inflection-point summaries, Welch-test comparisons, and mutual-information sensitivity rankings.

To preserve the anonymised review process, the uploaded model files and documentation have been anonymised, and the archive remains private during review. The archive will be made public and updated with full citation and author information upon article publication.

\section*{Acknowledgements}
\noindent The authors gratefully acknowledge Virginia Commonwealth University in Qatar and Qatar Foundation for supporting this work through the VCU Qatar Mathematical Data Sciences Lab.

\bibliographystyle{plain}

\bibliography{references}

\end{document}